\documentclass[preprint2]{aastex}

\newcommand{\solar}{$_{\odot}$}
\newcommand{\ceto}{C$^{18}$O}
\newcommand{\hcop}{HCO$^+$}
\newcommand{\httco}{H$^{13}$CO$^+$}
\newcommand{\nnh}{N$_2$H$^+$}
\newcommand{\httcn}{H$^{13}$CN}
\newcommand{\joz}{$J$=1$\rightarrow$0}
\newcommand{\jft}{$J$=4$\rightarrow$3}
\newcommand{\kms}{\,km\,s$^{-1}$}
\newcommand{\degree}{$^{\circ}$\hspace{-1mm}}
\def\lapp{\ifmmode\stackrel{<}{_{\sim}}\else$\stackrel{<}{_{\sim}}$\fi}
\def\gapp{\ifmmode\stackrel{>}{_{\sim}}\else$\stackrel{>}{_{\sim}}$\fi}

\shorttitle{Gravitational Infall in a Protostellar Cluster}
\shortauthors{Barnes et al.}

\begin{document}


\title{Discovery of Large-Scale Gravitational Infall \\
    in a Massive Protostellar Cluster}


\author{Peter J. Barnes\altaffilmark{1,2}, Yoshinori Yonekura\altaffilmark{3,4}, Stuart D. Ryder\altaffilmark{5}, Andrew M. Hopkins\altaffilmark{1,5},\\ Yosuke Miyamoto\altaffilmark{6}, Naoko Furukawa\altaffilmark{6}, and Yasuo Fukui\altaffilmark{6}}
\email{peterb@astro.ufl.edu}


\altaffiltext{1}{School of Physics A28, University of Sydney, NSW 2006, Australia}
\altaffiltext{2}{Astronomy Department, University of Florida, Gainesville, FL 32611, USA}
\altaffiltext{3}{Department of Physical Science, Osaka Prefecture University, 1-1 Gakuen-cho, Sakai, Osaka 599-8531, Japan}
\altaffiltext{4}{Center for Astronomy, Ibaraki University, 2-1-1 Bunkyo, Mito, Ibaraki 310-8512, Japan}
\altaffiltext{5}{Anglo-Australian Observatory, PO Box 296, Epping, NSW 1710, Australia}
\altaffiltext{6}{Department of Astrophysics, Nagoya University, Furo-cho, Chikusa-ku, Nagoya 464-8602, Japan}


\begin{abstract}
We report Mopra (ATNF), Anglo-Australian Telescope, and Atacama Submillimeter Telescope Experiment observations of a molecular clump in Carina, BYF73 = G286.21+0.17, which give evidence of large-scale gravitational infall in the dense gas.  From the millimetre and far-infrared data, the clump has mass $\sim$ 2$\times$10$^4$\,M\solar, luminosity $\sim$ 2--3$\times$10$^4$\,L\solar, and diameter $\sim$\,0.9 pc.  From radiative transfer modelling, we derive a mass infall rate $\sim$ 3.4$\times$10$^{-2}$ M\solar yr$^{-1}$.  If confirmed, this rate for gravitational infall in a molecular core or clump may be the highest yet seen.  The near-infrared $K$-band imaging shows an adjacent compact HII region and IR cluster surrounded by a shell-like photodissociation region showing H$_2$ emission.  At the molecular infall peak, the $K$ imaging also reveals a deeply embedded group of stars with associated H$_2$ emission.  The combination of these features is very unusual and we suggest they indicate the ongoing formation of a massive star cluster.  We discuss the implications of these data for competing theories of massive star formation.
\end{abstract}




\section{Introduction}

Many details of massive star formation in dense molecular clouds are still unclear \citep{Chu02}, despite much recent progress \cite[e.g.,][]{SBS02,FWS05,LBB07}.  For example, it is still debated whether massive stars can form by a scaled-up version of the accretion thought to occur with low-mass protostars \cite[e.g.][]{MT03}, or rather form by collective processes in a clustered environment \cite[e.g.][]{BBV03}.  Consequently, examples of massive star formation showing evidence of either behaviour can be informative to this debate, especially since there are still relatively few examples known of true massive protostars.

As part of the {\it Census of High- and Medium-mass Protostars} (CHaMP, \S\ref{strategy}), we identified the massive dense clump\footnote{Here we use the \citet{WBM00} terms for ``core''  (that part of a molecular cloud which will collapse to form an individual star or binary) and ``clump'' (which will collapse and fragment, via many cores, to form a star cluster).} G286.21+0.17 as showing striking evidence of large-scale gravitational infall, which we report here.  This source \cite[hereafter referred to as BYF73, from the master CHaMP source list;][]{BYF09} has been included in some previous surveys \citep{BNM96,DBS03,FBG04,YAK05} and is an Infrared Astronomical Satellite (IRAS) point source, but has not previously been shown to be remarkable.  The precise location is ($l$,$b$) = (286\degree.208, +0\degree.169) or ($\alpha$,$\delta$)$_{J2000}$ = ($10^{h}38^{m}32^{s}\hspace{-1mm}.2, -58^{\circ}19'12''$), about 1\degree.5 northwest of $\eta$ Carinae and 12$'$ north of the rim of the 15$'$-diameter HII region/bubble NGC\,3324/IC\,2599, at an assumed distance of 2.5 kpc.

\section{Observations}
\subsection{Survey Strategy \label{strategy}}

The motivation for CHaMP is to make a complete and unbiased census of higher-mass star formation at many different wavelengths over a large portion of the Milky Way \citep{BYM06}, in order to systematically characterise the processes in massive star formation in a uniform way.  The first step was to identify 209 dense clumps from \ceto\ and \hcop\ maps made with the 4m Nanten telescope \citep{YAK05,YBF09} of a $20^{\circ}\times6^{\circ}$ region of the Galactic Plane in Vela, Carina, and Centaurus (specifically $300^{\circ} > l > 280^{¡\circ}$ and $-4^{¡\circ} < b < +2^{¡\circ}$).  A higher-resolution follow-up campaign was then begun to map these clumps in a number of 3-millimetre wavelength (3mm) molecular transitions with the 22m-diameter Mopra dish of the Australia Telescope National Facility\footnote{The Mopra telescope is part of the Australia Telescope which is funded by the Commonwealth of Australia for operation as a National Facility managed by CSIRO.  The University of New South Wales Digital Filter Bank used for the observations with the Mopra telescope was provided with support from the Australian Research Council.} \citep{BYF09}.  The Mopra antenna's performance has been described by \citet{LPW05}.  Since that study, an on-the-fly (OTF) mapping capability has been implemented in the control software (in 2004), new 3mm MMIC receivers were installed (in 2005) which were at least as sensitive as the previous SIS mixers and much more efficient to operate, and the MOPS wideband digital filterbank was commissioned \cite[in 2006;][]{WMF06}.  These innovations, when combined with the Nanten maps as finder charts, makes an ambitious survey like CHaMP possible.

Mopra's OTF mapping mode has been described by T. Wong (2005, unpublished), which we briefly summarise here.  The telescope is driven in a raster pattern (which can be in any of the $l$, $b$, $\alpha$, or $\delta$ directions) at a rate such that the data dump interval (usually every 2\,s) from the correlator to mass storage is consistent with Nyquist- (or better) sampling of the sky, given the telescope beam and observing frequency.  At 90 GHz this drive rate across the sky equates to approximately $6''$\,sec$^{-1}$.  Each raster row is then offset by a similar amount (i.e.,\,$12''$ at 90\,GHz) from the previous row, until a square map with a size of the user's choosing is built up.  The user also selects whether a reference position (which can be specified in either relative or absolute coordinates) is observed at the beginning of each row, or only once every 2 rows.  Additionally, the user can choose from which corner of the square map the raster pattern is begun, i.e.\,the NE, NW, SE, or SW (in the respective coordinate system being used).  Finally the frequency of hot-cold load measurements of $T_{sys}$ needs to be specified; this is typically every 10-30\,min, depending on the stability of sky conditions.  In the 2007 season, however, a noise-diode calibration system was introduced into the data stream, effectively giving continuous $T_{sys}$ measurements and making separate hot-cold load scans somewhat redundant.  With the addition of a calibration spectrum of a known source such as Orion-KL, skydip measurements of the atmospheric opacity were not found to be necessary.

In this way a typical 5$'\times$5$'$ map can be built up over a period of about 70\,min at 90\,GHz.  In order to minimise rastering artifacts, however, a second map is usually made of the same field, but in an orthogonal rastering direction.  Including time ($\sim$10\,min) for pointing checks between each map, such a 5$'\times$5$'$ field is ``complete'' in about 2.5\,hr. Further rasters can be done of the same field, and this not only improves the S/N in the usual way, but under variable sky conditions will also minimise noise variations across a map, which might otherwise give erratic sensitivity coverage of the user's field.  After just 2 raster maps, however, the noise variations are usually acceptable ($\lapp$20\%) in all but the worst conditions.

The MOPS backend can be employed in either ``broadband'' or ``zoom'' mode.  With the former, the full 8\,GHz available bandwidth is observed with 65536 125-kHz-wide channels in each polarisation, corresponding to a velocity resolution of 0.45\kms\ at 90\,GHz.  In contrast the latter allows up to sixteen independently selectable 138-MHz-wide ``zoom IFs'' to be observed simultaneously from within the filterbank's 8\,GHz total instantaneous bandwidth.  Each zoom mode is correlated with 4096 channels in each of two orthogonal polarisations, resulting in a spectral resolution of 33\,kHz, or 0.11\kms\ at 90 GHz.  In the 2005--07 austral winter seasons we mapped the brightest 118 Nanten clumps, simultaneously covering many spectral lines in the 85--93 GHz range, among them the \joz\ transitions of \hcop, HCN, \nnh, \httco, and \httcn.  At these frequencies, Mopra has a beam FWHM of 36$''$, an inner error beam which extends to $\sim$80$''$, and a coupling efficiency of 0.64 to sources of this size \citep{LPW05}.

While CHaMP's 3mm molecular maps reveal the location of dense gas, complementary near-IR imaging of the same clumps can show where star formation has evolved further.  By compiling these statistics uniformly we will be in an excellent position to identify demographic trends in the massive star formation process.  Thus, an equally important part of CHaMP is a near-IR survey of the Nanten clumps using the IRIS2 imager \citep{TRE04} on the Anglo-Australian Telescope (AAT).  With this instrument we have begun acquiring images of each clump in $K$-continuum, Brackett-$\gamma$ (a recombination line tracing HII regions), and H$_2$ $v$=1$\rightarrow$0 $S$(1) \& $v$=2$\rightarrow$1 $S$(1) (vibrational quadrupole lines tracing molecular gas heated to a few 1000 K) to delineate the relationship between formed and forming massive stars, and report here results of such imaging.

\begin{figure*}[t]
\includegraphics[angle=-90,scale=0.61]{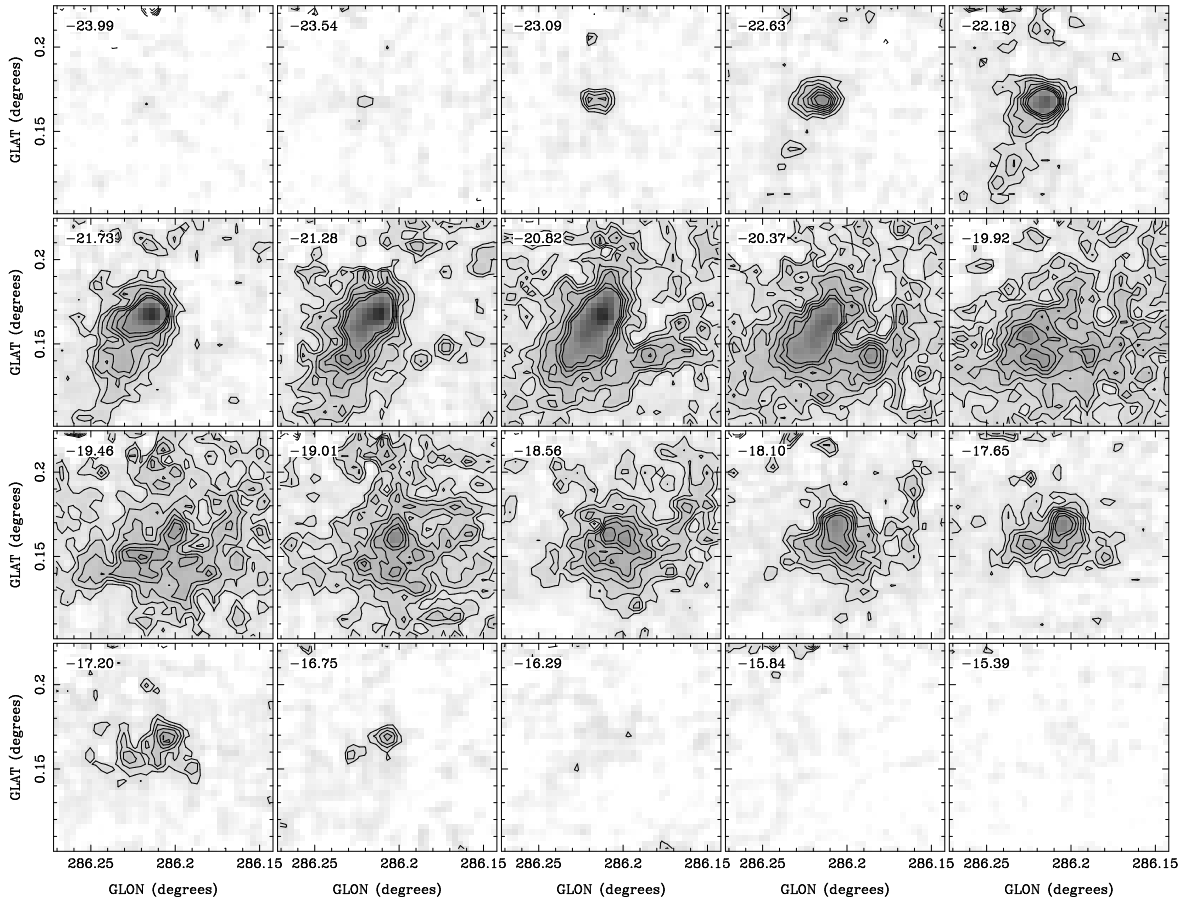}
\caption{Mopra channel maps of \hcop \joz\ emission from an inner 7$'$-field of BYF73 = G286.21+0.17, shown on the $T_A^*$ scale; $T_R^* = T_A^*/0.64$.  Each panel is actually the mean of four 0.113\,\kms\ wide channels, hence the panel spacing of 0.453\,\kms.  The grey scale is linear from 0\,K to the peak at 2.19\,K, while the contour levels are from 3 to 9 times the rms noise level for each panel, which is 0.092\,K.  At a distance of 2.5\,kpc, the scale is 40$''$ = 0.485\,pc or 0\degree.02 = 0.873\,pc or 1\,pc = 0\degree.0229 = 82$''$\hspace{-1mm}.5.
\label{chanmaps}}
\end{figure*}

A third major component of CHaMP will be a deep imaging survey of 1.2mm continuum and spectral-line emission with Atacama Submillimeter Telescope Experiment (ASTE), the 10m submillimeter telescope of the Nobeyama Radio Observatory\footnote{The ASTE project is led by Nobeyama Radio Observatory (NRO), a branch of National Astronomical Observatory of Japan (NAOJ), in collaboration with the University of Chile and Japanese institutes including the University of Tokyo, Nagoya University, Osaka Prefecture University, Ibaraki University, Kobe University, and Hokkaido University.} at Pampa la Bola in Chile \citep{Koh04,EKK04}.  The 1mm continuum is important in characterising the spectral energy distributions (SEDs) of embedded protostars as well as starless cores or clumps, and in correlating this with the phenomenology seen in spectral lines and at other wavelengths.  We report here as well some of the first data from this facility, namely \hcop\ and \httco\ $J$=4$\rightarrow$3 spectra, confirming the evidence of infall from our Mopra data.

\subsection{Observational Details and Data Reduction}

\begin{figure}[t]
\hspace{-10mm}
\includegraphics[angle=-90,scale=0.44]{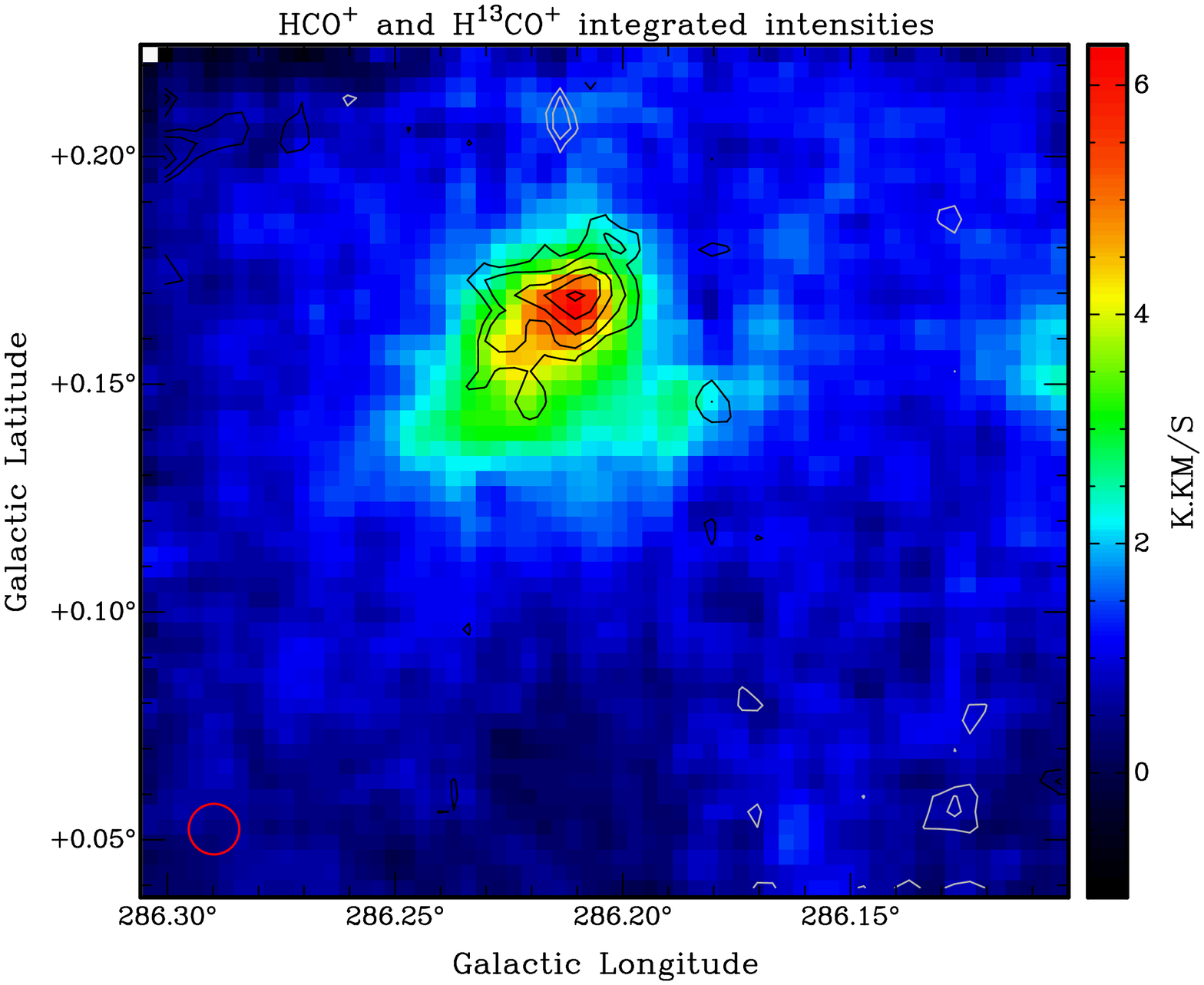}
\caption{({\it Image}) Wider 11$'$-field of Mopra \hcop \joz\ integrated intensity from BYF73, on the $T_A^*$ scale as given by the colourbar.  
The integration is over the range --23.20 to --16.63\,\kms\ or 58 channels, yielding an rms noise level 0.16\,K\,\kms: hence the widespread low-level emission above $\sim$0.5\,K\,\kms\ is real.  ({\it Contours}) Mopra \httco \joz\ integrated intensity in $T_A^*$, levels are ({\it grey}) --0.5, --0.35, ({\it black}) 0.35, 0.5, 0.7, 0.9, and 1.1\,K\,\kms.  The integration is from $-21.94$ to $-17.86$\,\kms, giving an rms noise level 0.12\,K\,\kms.  The smoothed Mopra HPBW for both datasets (40$''$) is shown for reference in the lower-left corner.
\label{mom0}}
\end{figure}

\begin{figure*}[t]
(a)\includegraphics[angle=-90,scale=0.34]{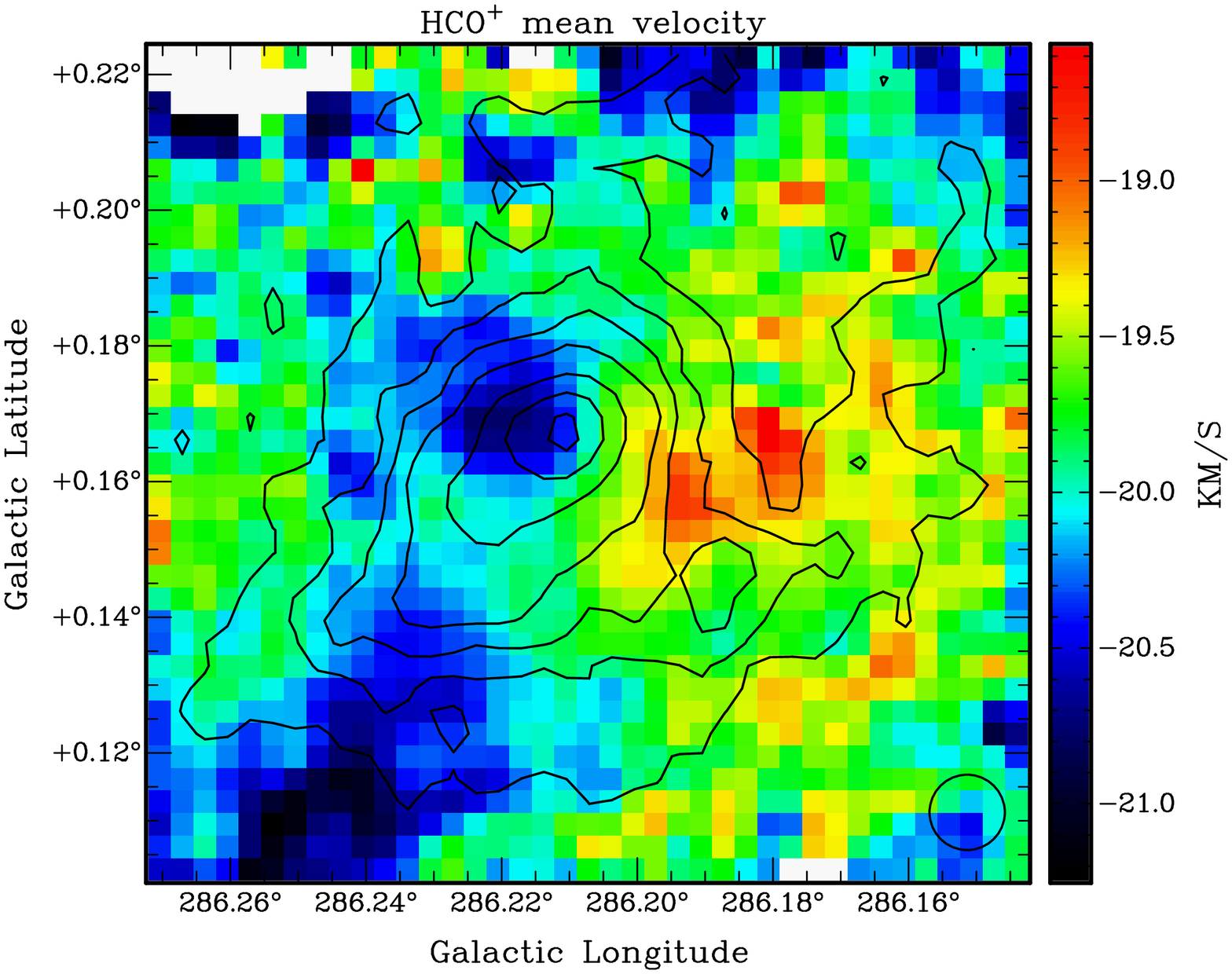}
(b)\includegraphics[angle=-90,scale=0.34]{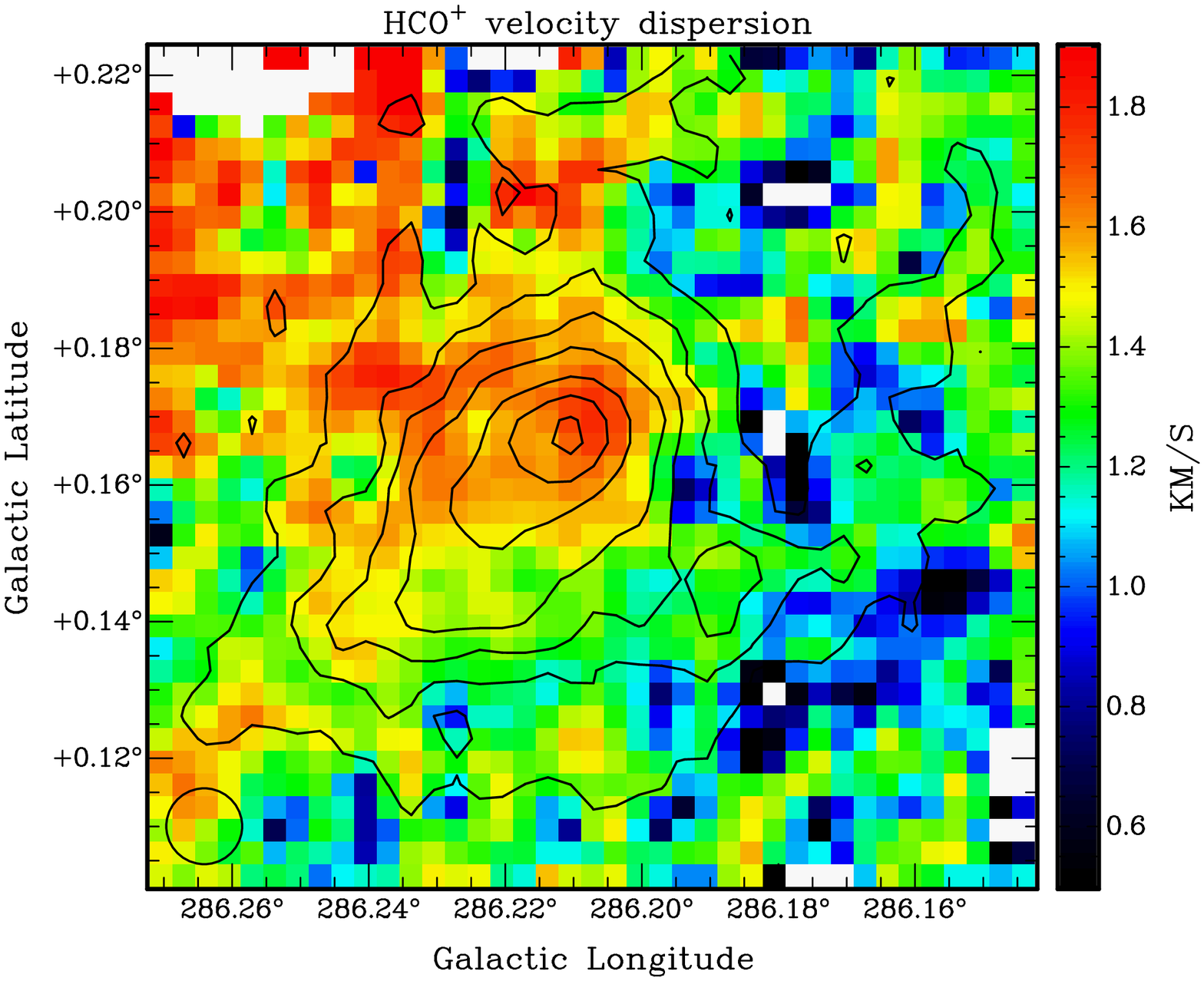}
\caption{Moment images of BYF73 from the Mopra \hcop\ data with telescope beam as shown in the corners, and overlaid with \hcop\ integrated intensity contours at 1.35, 1.9, 2.4, 3.0, 4.0, 5.0, and 6.0\,K\,\kms\ ($\sigma$ = 0.16\,K\,\kms). 
All moments for \hcop\ were calculated over the same velocity range as in Fig.\,\ref{mom0}({\it image}).  (a) First moment (intensity-weighted mean velocity field).  (b) Second moment (velocity dispersion).
\label{mom1mom2}}
\end{figure*}

The evidence for infall in BYF73 was first seen in the Mopra \hcop\ and \httco \joz\ data, presented in Figures \ref{chanmaps}--\ref{spectra}.  These were obtained on 2006 Oct 27--29 and 2007 Sep 5--9, when observing conditions were good ($T_{sys} \sim 300$ K or better).  The images were formed by coadding $5'\times5'$ OTF maps which abut each other to cover larger areas.  The reference position used for sky-subtraction during all mapping was ($l$,$b$) = (285.7,--0.3), which shows no emission in the Nanten CO map.  Each $5'\times5'$ area was scanned twice or three times in each of $l$ and $b$ in order to minimise rastering artifacts and noise variations.  The raw OTF data were processed with the Livedata-Gridzilla package \citep{BSD01} by bandpass division and baseline subtraction. The 2s-long OTF samples were then regridded onto a regular grid of 12$''$ pixels, where the samples were weighted by $T_{sys}^{-2}$, before averaging them into each gridded pixel.  Weighting by the rms$^{-2}$ of the spectra was not an option provided by Gridzilla; however as described above, since 2007 the continuously-measured $T_{sys}$ has effectively given the same information for each 2\,s sample.  For all Mopra maps in Figures \ref{chanmaps}--\ref{IRcolour}, the effective telescope HPBW has been smoothed at the gridding stage to 40$''$ from the intrinsic 36$''$, in order to reduce noise artifacts.  The resulting spectral line data cubes have low but, due to variations in weather and coverage, somewhat variable rms noise levels, ranging from 0.17\,K in the southern portions of Figure \ref{mom0} up to 0.22\,K in the north, per 0.11\kms\ channel; the average across the map is 0.2\,K per channel.  Although the pointing (checked on the SiO maser source R Carinae every hour or two) was typically good to 10$''$ or better ($<$1 pixel on the scale of our maps), because of the simultaneity of the spectral line mapping afforded by MOPS, the relative registration of features between these lines is perfect.

Observations of \hcop\ and \httco \jft\ were made using ASTE on 2006 Dec 1--2, when the typical system temperature (double-sideband) ranged from 220 K to 580 K at 345 GHz, including the atmosphere.  The half-power beamwidth of the telescope is 22$''$ at 345 GHz and the front end is a 4 K cooled SIS mixer receiver; at this frequency the beam efficiency is 0.65 \citep{Koh05,KNT08}.  We used a digital correlator with a bandwidth of 128 MHz and 1024 channels \citep{SSO00}.  The effective spectral resolution was 151.25\,kHz, corresponding to a velocity resolution of 0.13\kms\ at 345 GHz.  The data were obtained in position switched mode centred on  ($l$,$b$) = (286\degree.2071, +0\degree.1692); the ``off'' position used for sky-subtraction was (286\degree.0407, +0.4612), which is also devoid of Nanten CO emission.  Observations were made remotely from an ASTE operation room in San Pedro de Atacama, Chile, using a network observation system, N-COSMOS3, developed by NAOJ \citep{KET05}.  For \hcop\ and \httco, the total integration times of the spectra were 180\,s and 620\,s, and the rms noise levels 0.57 and 0.21\,K per channel, respectively.  The intensity was calibrated by using a room-temperature chopper wheel.  The absolute intensity was calibrated by observing Orion-KL and assuming $T_R^*$(\hcop) = 47\,K and $T_R^*$(\httco) = 2\,K \citep{SGB97}.  The pointing accuracy was measured to be reliable within 5$''$ as checked by optical observations of a star with a CCD camera attached to the telescope, as well as by CO $J$=3$\rightarrow$2 observations of IRC+10216.

IRIS2 observations were made at the AAT in service mode on 2006 May 13 in the $K$-band continuum (2.25--2.29 $\mu$m) and 3 spectral line filters as described above, Br-$\gamma$ and H$_2$ $S$(1) $v$=1$\rightarrow$0 \& $v$=2$\rightarrow$1.  For each filter, nine 60s images (dithered by 1$'$) were obtained of the instrument's 7$'$\hspace{-1mm}.7$\times$7$'$\hspace{-1mm}.7 field under 0$''$\hspace{-1mm}.9 seeing, and reduced using the ORAC--DR data reduction pipeline \citep{CHJ03}.  However the conditions during the observations were non-photometric, there being a fair amount of bushfire haze present.  Subsequent image processing was performed with the IRAF\footnote{IRAF is distributed by the National Optical Astronomy Observatories, which are operated by the Association of Universities for Research in Astronomy, Inc., under cooperative agreement with the U.S. National Science Foundation.} package.  The images for each field were registered using astrometry derived from SuperCOSMOS\footnote{This research has made use of data obtained from the SuperCOSMOS Science Archive, prepared and hosted by the Wide Field Astronomy Unit, Institute for Astronomy, University of Edinburgh, which is funded by the UK Science and Technology Facilities Council.} I-band images; we estimate the resulting rms positional accuracy in the IRIS2 images to be $<0''$\hspace{-1mm}.3.  Next, we linearly scaled the spectral-line images to the same relative brightness scale as the $K$-band continuum by matching the integrated fluxes of several stars in each filter, assuming they were of similar colour.  We then subtracted the continuum from the spectral-line images before transforming each image to Galactic coordinates.  Finally, a three-colour image (shown in Fig.\,\ref{IRcolour}a) was formed from the continuum-subtracted Br-$\gamma$ and H$_2$ images: Br-$\gamma$ is shown as red, and H$_2$ $S$(1) $v$=1$\rightarrow$0 \& $v$=2$\rightarrow$1 are shown as green \& blue, respectively.

Long-slit spectroscopy with IRIS2 was obtained on 2007 Oct 18.  The 7$'$\hspace{-1mm}.7 long slit was set to a position angle of 131\degree.2, with the stellar cluster and nebulosity of BYF73 spanning most of one half of the slit.  Four exposures of 300s were obtained in the $K$-band, with BYF73 nodded by 3$'$\hspace{-1mm}.8 along the slit between each exposure.  Similar nodded exposures of the nearby A0V star HD 95534 were obtained to assist with telluric correction.  All frames were flatfielded using quartz lamp exposures, then nodded pairs were subtracted to remove sky emission.  After two-dimensional wavelength calibration and straightening with Xe lamp exposures, the ``off'' beam data were inverted, aligned, and co-added to the ``on'' beam data.  Each spectral row of the data was divided by an extracted spectrum of HD 95534 (from which intrinsic Br-$\gamma$ absorption had been removed), then multiplied by a blackbody spectrum of $T_{eff}$ = 9520 K.

\notetoeditor{Figures 3a and 3b should appear side-by-side in print} 

\section{Analysis and Discussion}
\subsection{Mopra Maps \label{mopra}}

The Mopra \hcop \joz\ maps, being of high signal-to-noise ratio, reveal a number of interesting features which we describe here.  Figure \ref{chanmaps} shows the \hcop\ emission from BYF73 across its full velocity range, where we have averaged four velocity channels into each displayed panel for ease of viewing.  However all analysis below rests upon the full-resolution data.  At the central velocities (--22 to --18\kms) the emission is quite widespread; redward of the line centre (--20 to --18\kms) this extended emission is quite clumpy, while to the blue (--22 to --20\kms) the emission is strongly centrally concentrated.  At both the reddest and bluest velocities, the emission is fairly centrally concentrated; in particular there is little obvious evidence for an extended, high-velocity outflow which would tend to have emission well away from the centre at the highest relative velocities.  In fact to a casual inspection there seems to be little systematic kinematic structure to these channel maps at all, and it is only upon inspection of the spectra that the infall profiles are revealed.

This is also reflected in the integrated intensity image of Figure \ref{mom0}, where the extended envelope of BYF73 shows very little evidence of being structurally disturbed by (for example) its proximity on the sky to $\eta$ Carinae or to NGC 3324.  The only morphological feature of note in the envelope is a bay to the NW, which as we shall see is intrinsic to the source.  The inner $2'$ of BYF73 also appears fairly bland: this area is elongated somewhat in the NW-SE direction in both the \hcop\ and \httco\ emission, and there is a small but significant offset in the peak positions of these two molecules, with the \httco\ emission centred slightly to the northwest of the brightest \hcop.

In Figure \ref{mom1mom2} we give the higher-order \hcop\ moment images, overlaid by the moment-0 contours from Figure \ref{mom0}.  In contrast to the latter, the intensity-weighted mean velocity (Fig.\,\ref{mom1mom2}$a$) reveals a striking velocity gradient across the clump; the axis of this gradient is rotated $\sim$\,30\degree\ anticlockwise from the long axis of the clump, as seen in the moment-0 contours.  The spectral line is most strongly blue-shifted to the north and east of the peak \hcop\ emission, reaching its minimum value $\sim$0$'$\hspace{-1mm}.5 east of the peak.  This blueshift gradually changes to a redshift to the western side of the clump, reaching its maximum value $\sim$2$'$ west of the peak, inside the bay of the envelope where the emission is weaker.  The weaker emission which wraps around the western side of the bay (the ``western arm'') is also reshifted with respect to the clump.  As a whole the clump's velocity is significantly blueshifted with respect to the \httco\ line centre (see below), which is approximately at the green colour in this image.

The \hcop\ velocity dispersion $\sigma$ (where the line FWHM = 2.355$\sigma$) is shown in Figure \ref{mom1mom2}$b$.  [Note that because the line shape is strongly non-Gaussian, one should not confuse this moment-2 measurement with an actual linewidth; nevertheless its variation does indicate true changes in the line profile across the source.]  Here again we see a strong gradient in this parameter: the bulk of the clump, and to its north and east down to an integrated intensity level $\sim$\,2.0\,K\kms, has a large $\sigma \sim$\,1.5--1.8\kms.  Below this intensity to the west and SW, $\sigma$ drops to 1.1--1.4\kms, reaching minimum values $<$1\kms\ exactly where the velocity field is most redshifted.  One therefore suspects that these redshifted features are due mostly to individual, narrower-line substructures in the envelope, and that apart from these features, the clump's overall blueshift with respect to the optically thin line centre is even more complete.

These images can be compared to the equivalent moment images of the \httco\ cube (not shown here, but see Fig.\,\ref{spectra}$b$ for a spectrum).  From Gaussian fits to this cube, the peak $T_A^* \sim$\,0.3--0.4\,K, and the line is centred near $V_{LSR} \sim$ --20 to --19.5\kms\ across most of the emission, but shifts to --19 to --18.5\kms\ along the clump's SW edge.  The line FWHM varies from $\sim$\,1--3\kms\ toward the SE end of the clump, rising to $\sim$\,2--5\kms\ toward its NW end.  Such fits to the \httco\ have rms residuals $\sim$\,0.17\,K per 0.12\kms\ channel, with typical uncertainties $\sim$\,0.3 and 0.7\kms\ to the $V_{LSR}$ and linewidth, respectively.

\subsection{Distance Determinations \label{distance}}

Typical Galactic rotation curves \citep{BG78,Cle85} and standard values of $R_0$ and $\Theta_0$ \cite[8.5\,kpc/220\,\kms, IAU 1978 values; 8\,kpc/200\,\kms, Merrifield 1992; 8.4\,kpc/254\,\kms,][]{Rei09} indicate that the central $V_{LSR}$ = --19.7\,\kms\ for the molecular clump (see \S\ref{radxfer}) is formally forbidden at the longitude of BYF73 (meaning that this velocity is inconsistent with such rotation curves for objects at {\em any} distance along this line of sight).  In fact the minimum allowed velocity for the tangent point at this longitude is $\sim -10$\kms, which is $\sim$2$\sigma$ more positive than $V_{BYF73}$ \cite[where $\sigma=4$\,\kms\ is the cloud-to-cloud velocity dispersion of GMCs;][]{BG78}.  In spite of this disparity, any other location for BYF73 is even less kinematically favoured than the tangent-point distance.  For example, BYF73 and the $\eta$ Car GMC may be sharing in a non-circular streaming motion of the order of 10\,\kms\ associated with this part of the Carina Arm.  
In any case, at such tangent points small uncertainties in the rotation curve or the values of $R_0$ and $\Theta_0$ can translate into large line-of-sight distance uncertainties, up to 50\% or more.  Therefore while a tangent-point distance of $R_0$cos$(l)$=2.35$\pm$1.5\,kpc \cite[using][]{Rei09} is favoured with the kinematic method, a more robust determination is preferred for the analysis in the sections following, especially in light of the large power with which the distance to BYF73 enters some of the formulae below.

Fortunately, a number of studies have yielded distances to the massive clusters in and near the $\eta$ Car GMC \cite[e.g., see the summary by][]{YAK05}, and to NGC\,3324 \citep{HCS78}.  These range from 2.2--2.8\,kpc, reassuringly close to the tangent-point distance.  Therefore if we adopt a mean value of $d$=2.5\,kpc we would likely only need to attach a 12\% uncertainty to it.  Although their association in velocity and on the sky is strong circumstantial evidence, it is not certain, however, that BYF73 is actually associated with the $\eta$ Car GMC complex or NGC\,3324.  In particular, Fig.\,\ref{mom0} shows no evidence that the low-density molecular envelope of BYF73 has been at all disturbed by the vigorous star-formation activity closer to $\eta$ Car or by the bubble of NGC\,3324.  Nevertheless, further evidence that the tangent-point distance is reasonable follows from analysis of the cm-continuum emission of the small HII region adjacent to BYF73 (see Fig.\,\ref{IRcolour} and \S\ref{IRfeatures}).  Using the MGPS-2 \citep{MMG07} and SGPS \citep{HGM06} flux densities at 843 and 1420 MHz of 62$\pm$5 and 85$\pm$11 mJy respectively, and assuming an electron temperature in the HII region $T_e=7000\pm1000$\,K \citep{SMN83}, standard analysis \citep{MST67,Bar85} gives a distance-independent emission measure EM$=(1.4\pm0.4)\times10^6$\,pc\,cm$^{-6}$, typical of compact HII regions \citep{HI79}.  Such HII regions have diameters $\sim$0.1--1\,pc, bracketing that for BYF73 (from measurement of the Br-$\gamma$ nebula in Fig.\,\ref{IRcolour}$a$, its FWHM = 0.25\,pc), and so yielding a most likely location for it at the tangent point.

In summary, various lines of reasoning make a good case for BYF73 lying close to the tangent-point distance for its longitude.  We therefore assign a distance of 2.5$\pm$0.3\,kpc, based on the direct measurements listed by \citet{YAK05}.

\subsection{Evidence for Gravitational Infall \label{gravinfall}}

\begin{figure*}[t]
\footnotesize{
\centerline{
(a)\includegraphics[angle=-90,scale=.262]{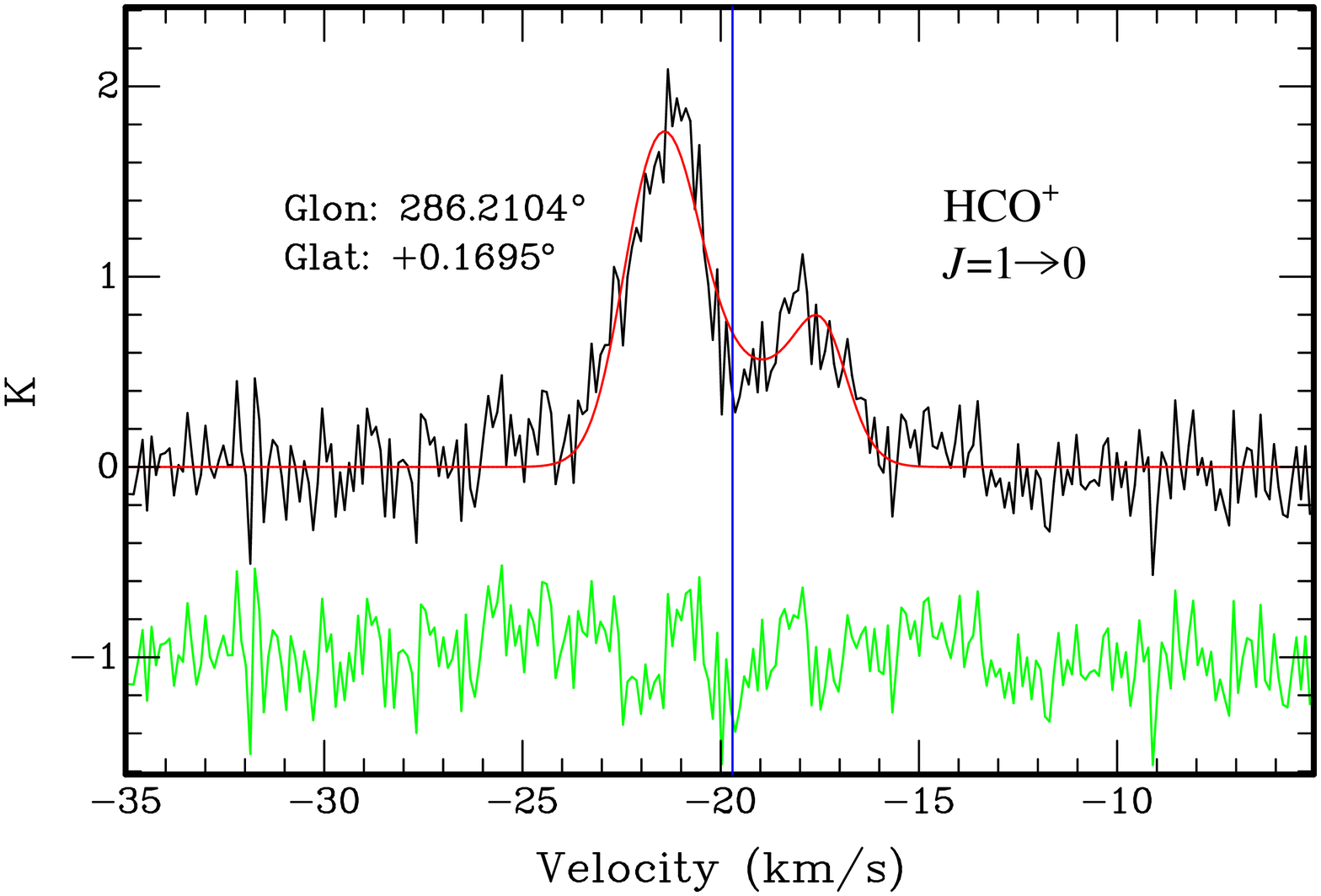}
(c)\includegraphics[angle=-90,scale=.258]{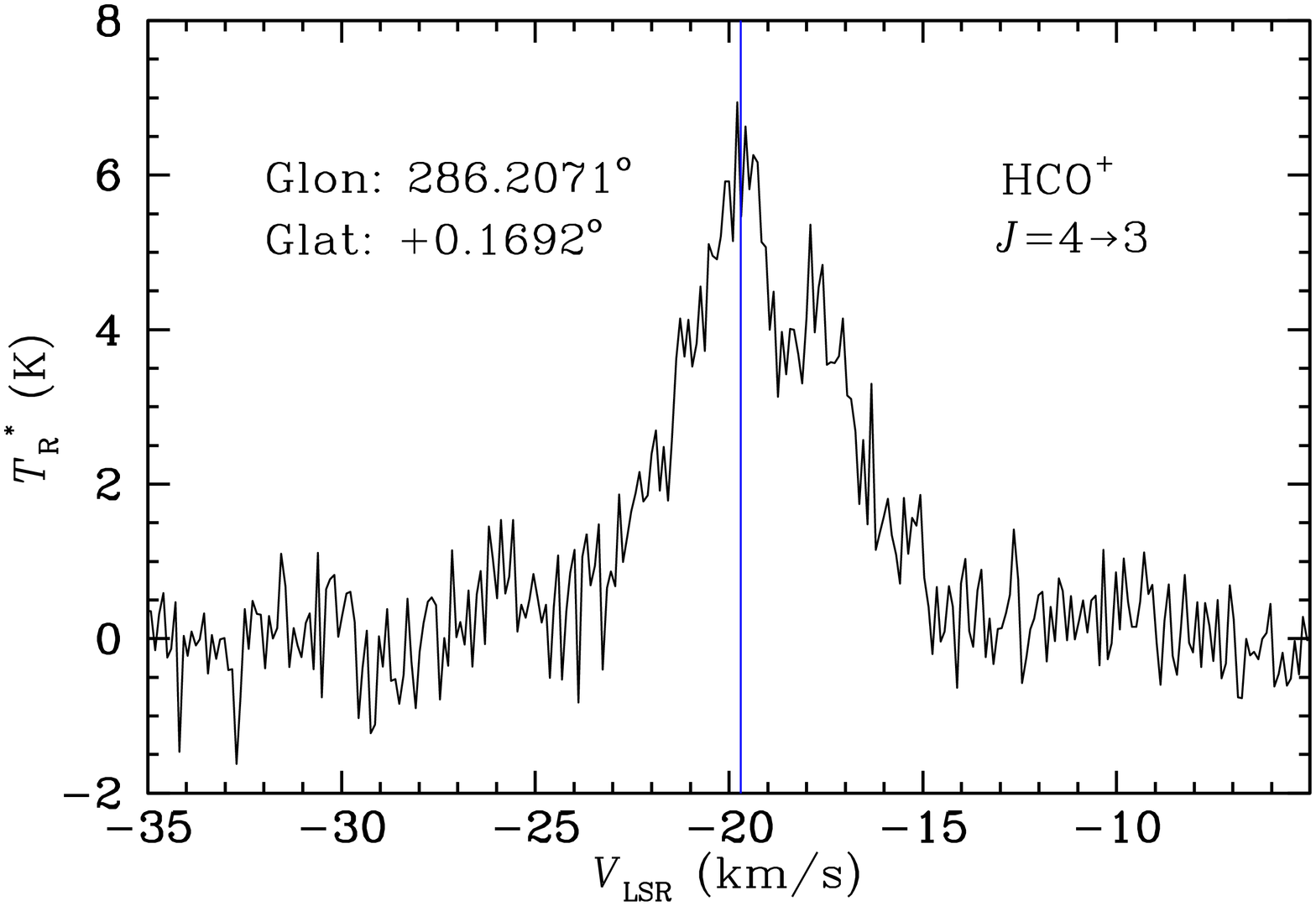}}
\centerline{
(b)\includegraphics[angle=-90,scale=.262]{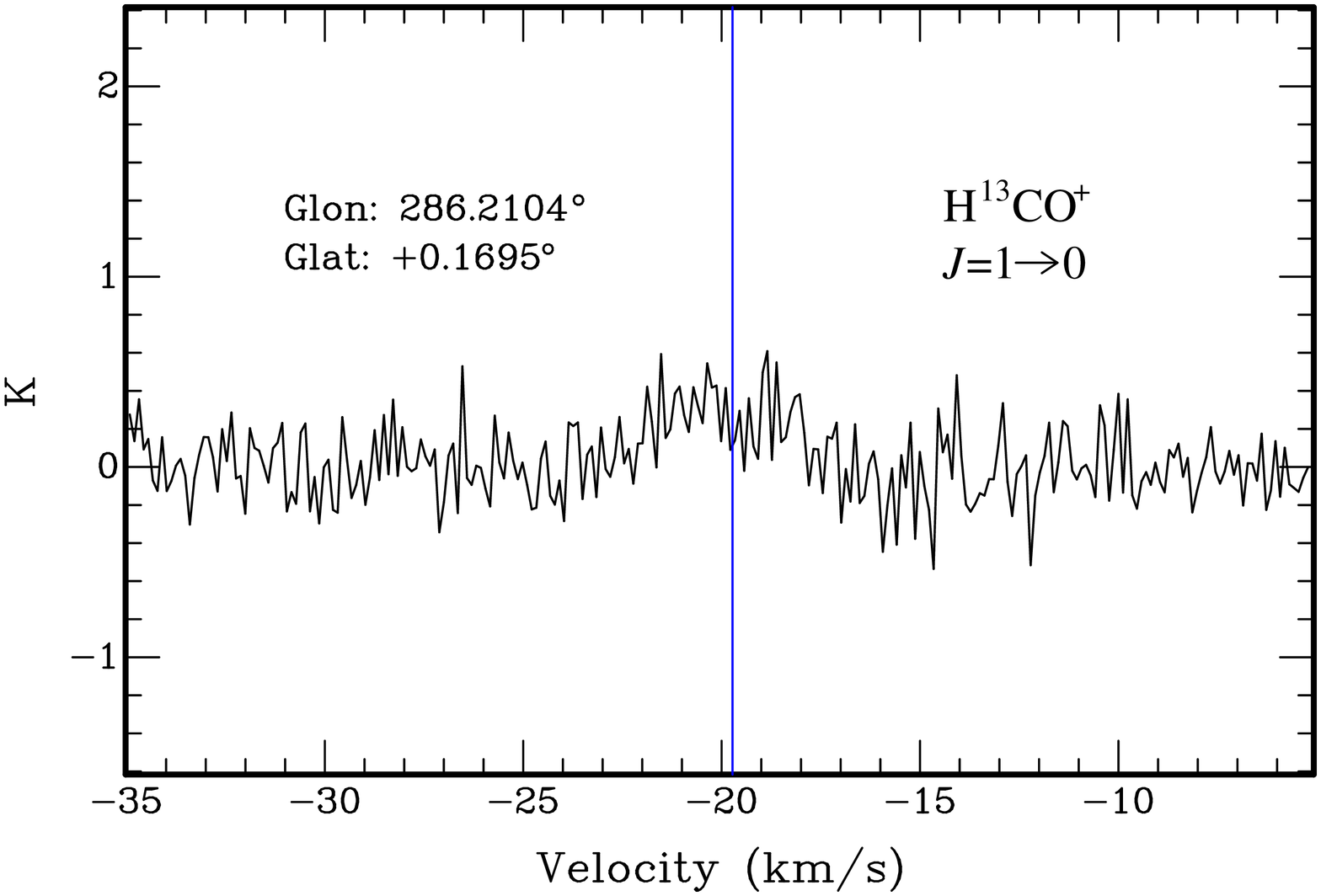}
(d)\includegraphics[angle=-90,scale=.258]{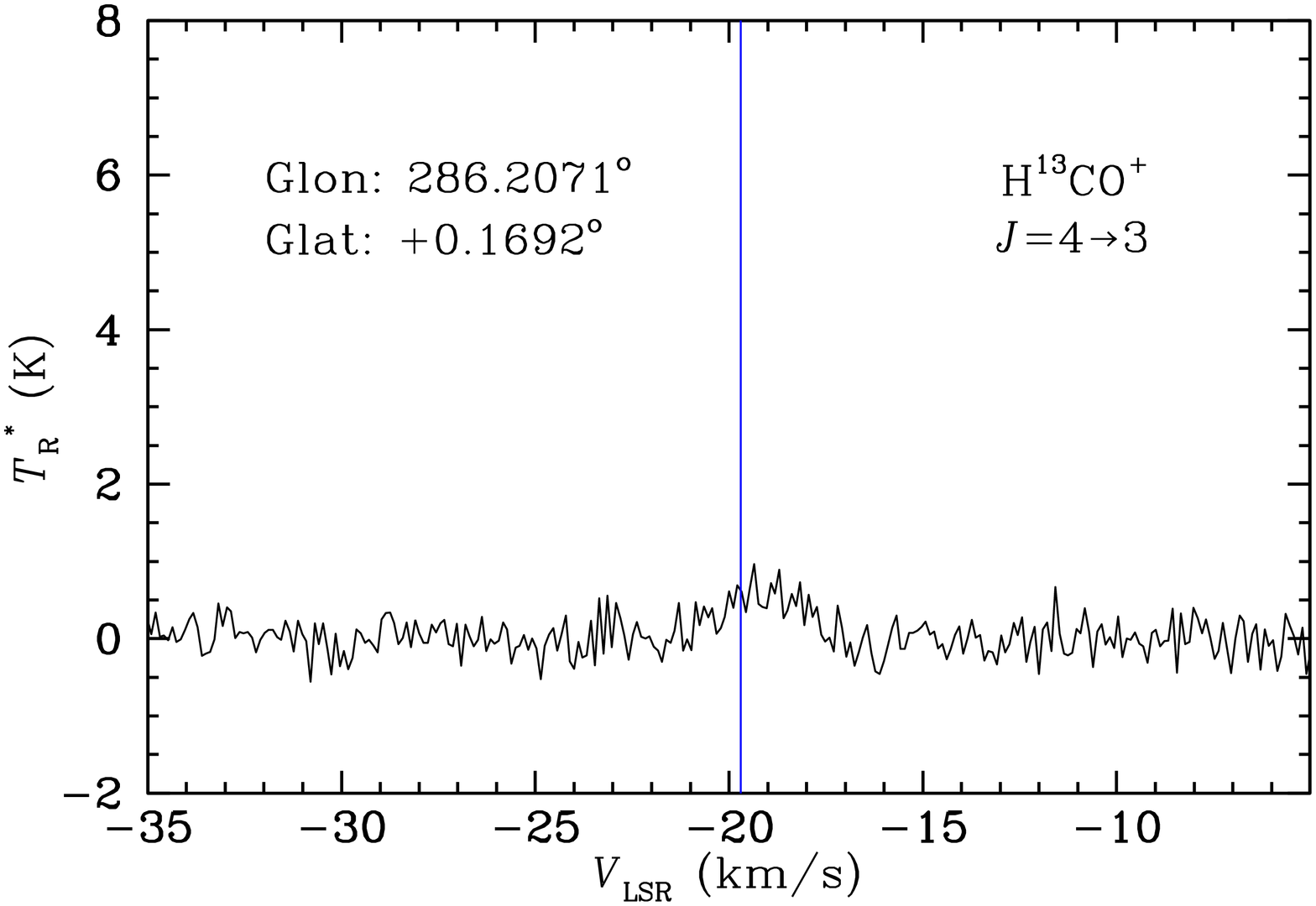}}
}
\caption{Sample Mopra and ASTE spectra of BYF73.  Panel {\it a} shows the \hcop \joz\ spectrum in black at the peak \httco\ position.  The Hill5 model fit at this point (see \S\ref{radxfer}) is shown in red, and the residual spectrum (data-model) is shown in green, offset 1K below the \hcop\ spectrum.  The vertical blue line indicates the systemic velocity at $V_{LSR} = -19.7$\kms, also from the model fits.  Panel {\it b} is the Mopra \httco \joz\ spectrum, also at the peak \httco\ position of BYF73, shown at the same $T_A^*$ and $V_{LSR}$ scales as panel {\it a}.  
Panels {\it c} and {\it d} are ASTE \jft\ spectra from the BYF73 peak position in \hcop\ and \httco, respectively, shown at the same velocity scale as panels $a$ and $b$, but on the $T_R^*$ scale.
\label{spectra}}
\end{figure*}

\begin{figure}[t]
\footnotesize{\centerline{
\includegraphics[angle=-90,scale=.322]{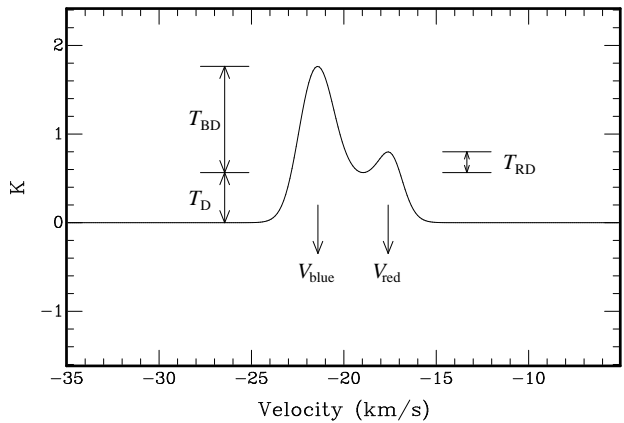}
}}
\caption{Sample Hill5 model spectrum (the same as Fig.\,\ref{spectra}$a$ in red) with measurements as shown following Fig.\,2$c$ of \citet{MMT96}.  
\label{model}}
\end{figure}

The dense molecular clump, centred at ($l$,$b$) = (286\degree.208,+0\degree.169) and easily visible in the Mopra maps, has \hcop\ spectral line profiles that fit the canonical pattern of \citet{ZEK93} indicating gravitational infall onto a protostar (see Fig.\,\ref{spectra}).  For the optically thick \hcop\ emission, this includes a self-absorbed profile with predominantly stronger blue wings at most positions (the ``blue asymmetry'' or inverse P-Cygni profile, seen in panels {\it a} and {\it c} of Fig.\,\ref{spectra}), together with more Gaussian line profiles for the optically thin transitions of \httco, which are centred in velocity on the \hcop\ self-absorption (panels {\it b} and {\it d}).  Further, the \jft\ lines (panels {\it c} and {\it d}) are brighter than the corresponding \joz\ lines (panels {\it a} and {\it b}), the self-absorption in the \hcop\ is deeper in the \joz\ than the \jft, the velocity difference $V_{blue} - V_{red}$ (see below) between the blue and red peaks of the \hcop\ lines is slightly greater in the \joz\ than the \jft, and the blue and red peaks in the \jft\ line are both slightly redward of the respective peaks in the \joz\ line.  All of these details are completely consistent with the \citet{ZEK93} and \citet{MMT96} picture of a dense core undergoing gravitational infall, where the velocity of the infall and the temperature both increase towards the centre, producing the respective line profiles and ratios.  

\notetoeditor{Figures 4a -- 4d should be large enough to be legible, probably taking up most of a page for the 4 panels.  Care should be taken to make the panels the same size and well-aligned with each other.}

However, the mass scale of the infall appears to be unusual.  \citet{MMT96} developed a simple but useful two-layer model to evaluate basic parameters from spectra of molecular cores which are undergoing gravitational infall, while \citet{DM05} extended this analytic model and provide a general code for robustly determining these parameters and their uncertainties.  Although these models were developed in the context of low-mass protostars, the results we derive here satisfy the assumptions made in their treatment of the radiative transfer.  The key qualifications are that the infall speed not be much greater, nor much smaller, than the velocity dispersion in the dense gas, which result we obtain below.  Here we use the \citet{MMT96} formalism and the \hcop\ line profiles to estimate the characteristic gas infall speed and motivate further discussion.  In \S\ref{radxfer} we use the \citet{DM05} code to more rigorously evaluate the model fits.

From \citet{MMT96}'s eq.\,(9) and using the parameters as listed in Table \ref{params} from the sample spectra in Figure \ref{spectra}, 
we obtain $V_{in}$ as shown also in Table \ref{params}, and where the quoted errors are obtained by propagating the measured uncertainties through the formula. 
Continuing to follow Myers et al., we need an estimate  for the radius over which the infall profile is seen, in order to allow calculation of a kinematic mass infall rate.  This profile is widespread in the \hcop\ data, but its intensity drops only slowly into the background, showing no hard edge.  To indicate a radius we consider the emission FWHMs (suitably deconvolved).  In the \hcop\ and \httco \joz\ data, the diameters $D_{FWHM} = 120''\pm4''$ and 
$65''\pm5''$ respectively, taking a geometric mean of the major and minor axes in each case.  At 2.5\,kpc these respectively give clump radii $R = 0.73\pm0.09$\,pc and $0.40\pm0.05$\,pc, where we have now also added in quadrature the uncertainty due to the distance.  Although the \hcop\ infall profiles are clearly more widespread than the \httco\ radius, we conservatively take the latter as an optically thin tracer and therefore more representative of the true column distribution, understanding that this may in fact be a lower limit to the clump radius.  This gives
\begin{eqnarray} 
	\frac{dM_k}{dt} & = & 4\pi R^2\mu_{mol} m_{H}n_{cr}V_{in} \nonumber \\
	 & \sim & (2.9\pm1.5)\times10^{-2}\,{\rm M}_{\odot}{\rm yr}^{-1} 
\end{eqnarray}

\hspace{-4mm}for BYF73's mass infall rate, where $\mu_{mol}$ = 2.30 is the mean molecular mass in the gas and $n_{cr}$ is the critical density for the \joz\ transition (see \S\ref{mass}).  This should be compared to the gravitational mass infall rate for the self-similar singular isothermal sphere (SIS) solution \citep{Shu77}
\begin{equation} 
	\frac{dM_g}{dt} = \frac{\sigma^3}{G} \sim 0.080\times10^{-2}\,{\rm M}_{\odot}{\rm yr}^{-1} , 
\end{equation}

\hspace{-4mm}where instead of the sound speed $c$ of Shu, we have substituted, as suggested by \citet{BP07}, the supersonic velocity indicated by the \httco\ linewidth $\sigma$ from Table \ref{params} (see also \S\ref{radxfer}).  Even so, we see that for BYF73, Shu's solution cannot give us the observed infall rate.  Instead, Banerjee \& Pudritz show that a magnetised core can collapse supersonically with an effective speed $Mc$, where $M$ is the Mach number in the flow.  For BYF73, then, the observed infall only requires collapse with $M \sim$ 3.  This Mach number and infall speed above are consistent with (for example) the MHD simulations of \citet{BP07} or the \citet{MT03} massive turbulent core model; however our mass infall rate is still more than an order of magnitude higher than in such models, mainly because of the large extent of the infall asymmetry in our maps.

\begin{table}[t]
\caption{Sample infall fitting parameters for the \citet{MMT96} model at the peak \httco \joz\ position}
\vspace{-3mm}
\begin{center}
{\small
\begin{tabular}{cccc}
  \hline\vspace{-2mm} \\
  Parameter & \joz & \jft & Units\vspace{1mm} \\
  \hline\vspace{-2mm} \\
  $V_{blue}$ & $-21.0\pm0.1^a$ & $-19.7\pm0.1^c$ & \kms \\
  $V_{red}$ & $-18.0\pm0.1^a$ & $-17.8\pm0.1^c$ & \kms \\
  $T_{D}$ & $0.44\pm0.02^a$ & $3.6\pm0.1^c$ & K \\
  $T_{BD}$ & $1.10\pm0.04^a$ & $2.6\pm0.2^c$ & K \\
  $T_{RD}$ & $0.24\pm0.04^a$ & $0.9\pm0.2^c$ & K \\
  $\sigma$ & $1.5\pm0.3^b$ & $1.06\pm0.15^d$ & \kms\vspace{1mm} \\
  \hline\vspace{-2mm} \\
  $V_{in}$ & $0.86\pm0.36$ & $0.34\pm0.12$ & \kms\vspace{1mm} \\
  \hline \\
\end{tabular} \\
}
\vspace{-2mm}
{\footnotesize
\textsc{Footnotes:} Measurements from (a) Fig.\,\ref{spectra}$a$; (b) Gaussian fitting (not shown) to Fig.\,\ref{spectra}$b$; (c) Fig.\,\ref{spectra}$c$; (d) Gaussian fitting (not shown) to Fig.\,\ref{spectra}$d$. \\
}
\end{center}
\label{params}
\vspace{-5mm}
\end{table}

\citet{MMT96} suggested that, for their low-mass protostars, the agreement of the inferred and theoretical rates indicates the derived inward motions are consistent with gravitational infall.  Under this interpretation BYF73 also gives a much larger infall rate than is typical of low-mass protostars \cite[$\sim$10$^{-6}$\,M\solar\,yr$^{-1}$, increasing to 10$^{-4}$ to 10$^{-5}$\,M\solar\,yr$^{-1}$ during FU Orionis-type outbursts;][]{Lad99}, again stemming mainly from the parsec-scale extent of the asymmetric \hcop\ profile, and also from the unusually large value for $V_{in}$.  This mass infall rate is also larger than any seen so far in any similar massive star-forming region \cite[e.g.][]{FWS05,BCC06}.  Given the linear size of this region and the near-IR appearance of peculiar emission-line nebulosity at the centre of the clump, we suspect that the entire BYF73 cloud is undergoing a global gravitational collapse.  Verification of this suggestion awaits additional supporting evidence including interferometric observations and more detailed modelling.  However all of the Mopra spectral line maps of BYF73 (e.g.\,HCN, \nnh, etc.), as well as the CS $J$=2$\rightarrow$1 data reported by \citet{BNM96}, show similar emission distributions and/or line profiles, with differences as expected from the species' different relative abundance.  This is not surprising considering that they all require high densities \cite[$n_{cr}\sim10^{5-6}$\,cm$^{-3}$;][]{Spi78} to be collisionally excited and thermalised to the gas kinetic temperature, and so should reflect the same dynamical state as seen in the \hcop.

\subsection{Radiative Transfer Modelling \label{radxfer}}

\citet{DM05} compared a number of analytic radiative transfer models of infall in a low-mass dense core to a full Monte Carlo model.  They found that their ``Hill5'' model gave the most accurate simulation of the Monte Carlo solution, and of all the analytic models they examined, was the most robust against various numerical and instrumental uncertainties.  We have used their HILL5 code to analyse our Mopra \hcop \joz\ data cube pixel-by-pixel, and present the results here.

\notetoeditor{The panels+caption of this figure should fill a page for clarity, eg in 3 rows of 2 panels each: a-b/c-d/e-f.  I couldn't get the latex to do this properly, so they are displayed here as 2 rows of 3 panels each.}
\begin{figure*}[t]
\centerline{
(a)\includegraphics[angle=-90,scale=0.21]{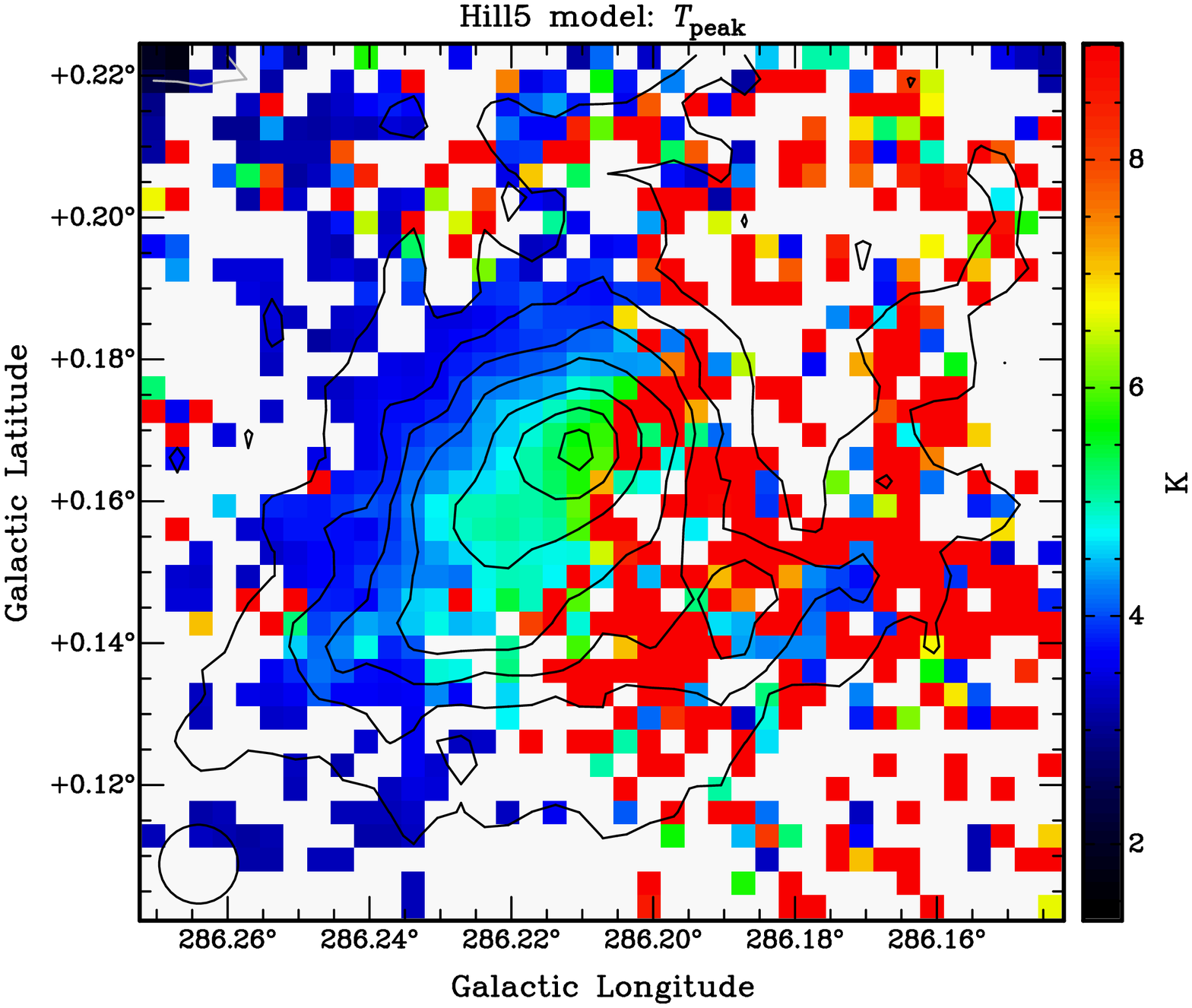}
(b)\includegraphics[angle=-90,scale=0.21]{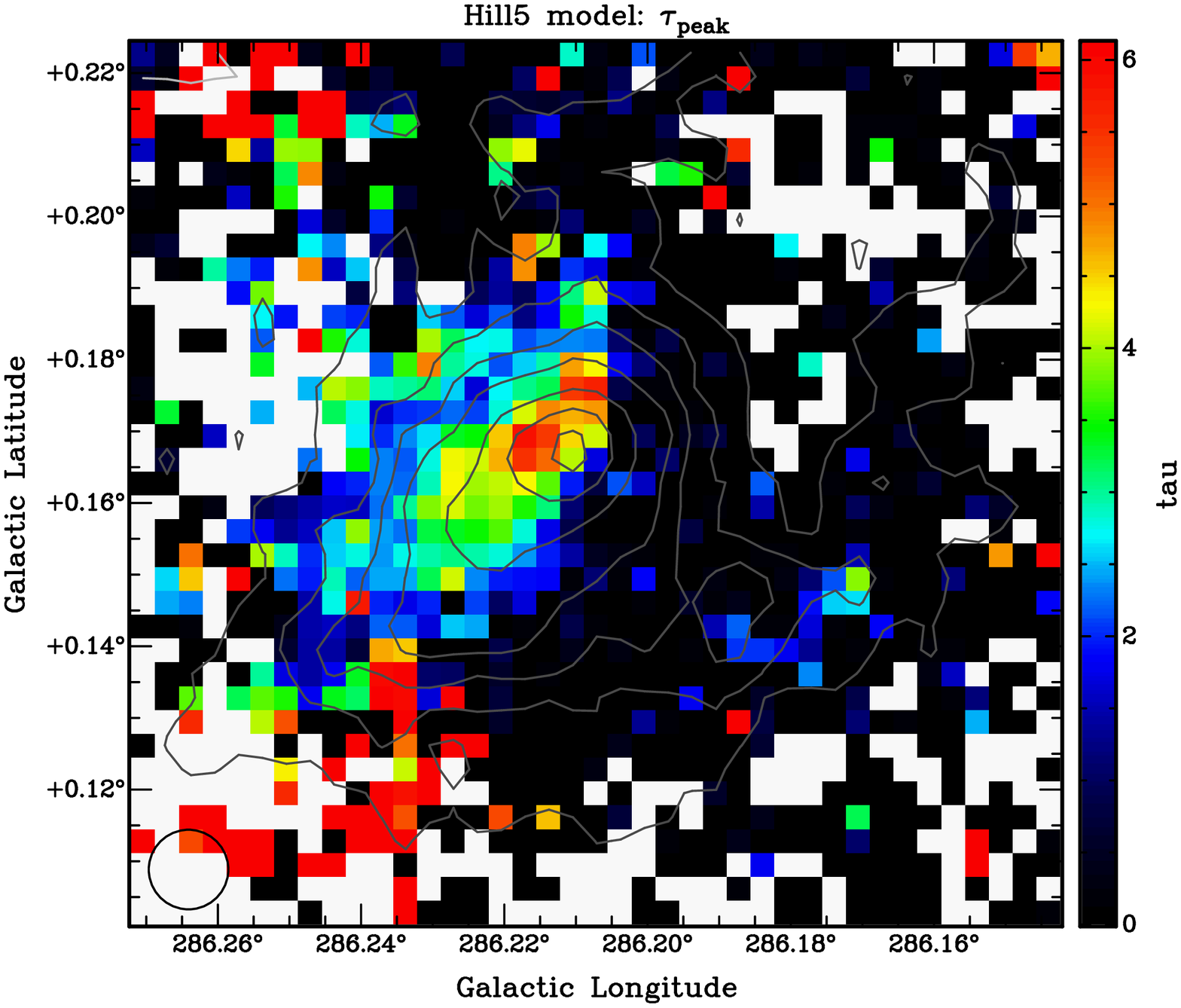}
(c)\includegraphics[angle=-90,scale=0.21]{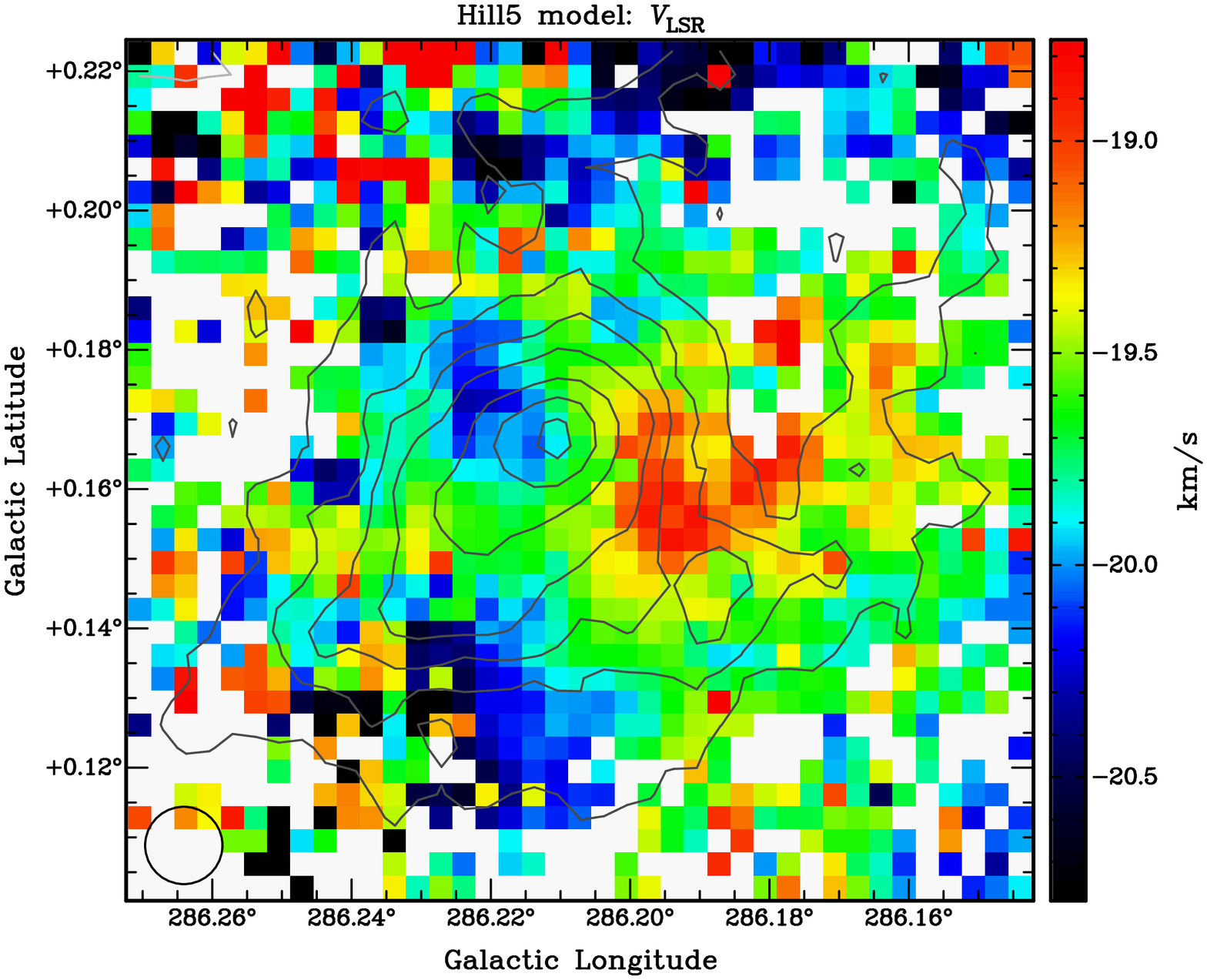}}
\centerline{
(d)\includegraphics[angle=-90,scale=0.21]{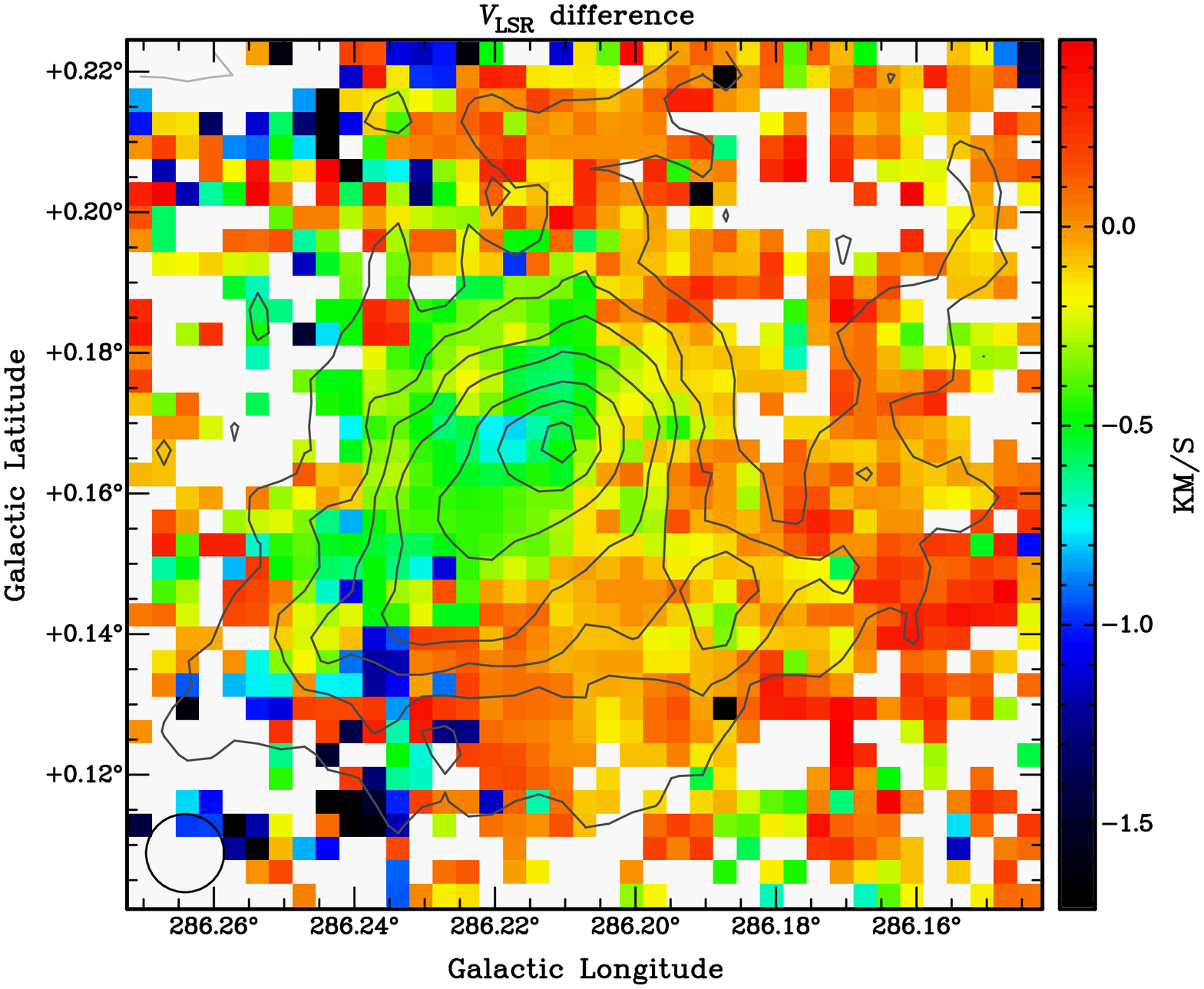}
(e)\includegraphics[angle=-90,scale=0.21]{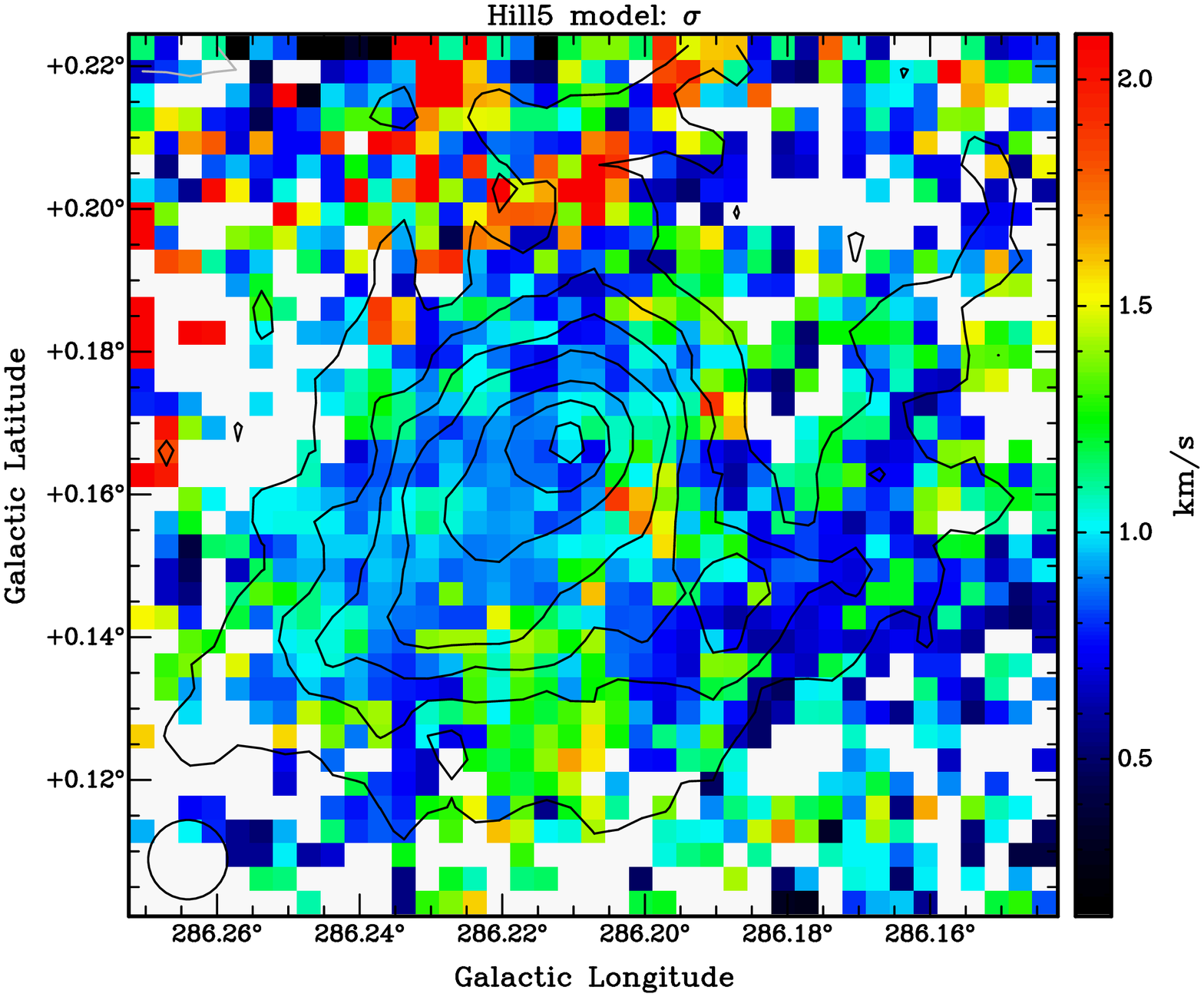}
(f)\includegraphics[angle=-90,scale=0.21]{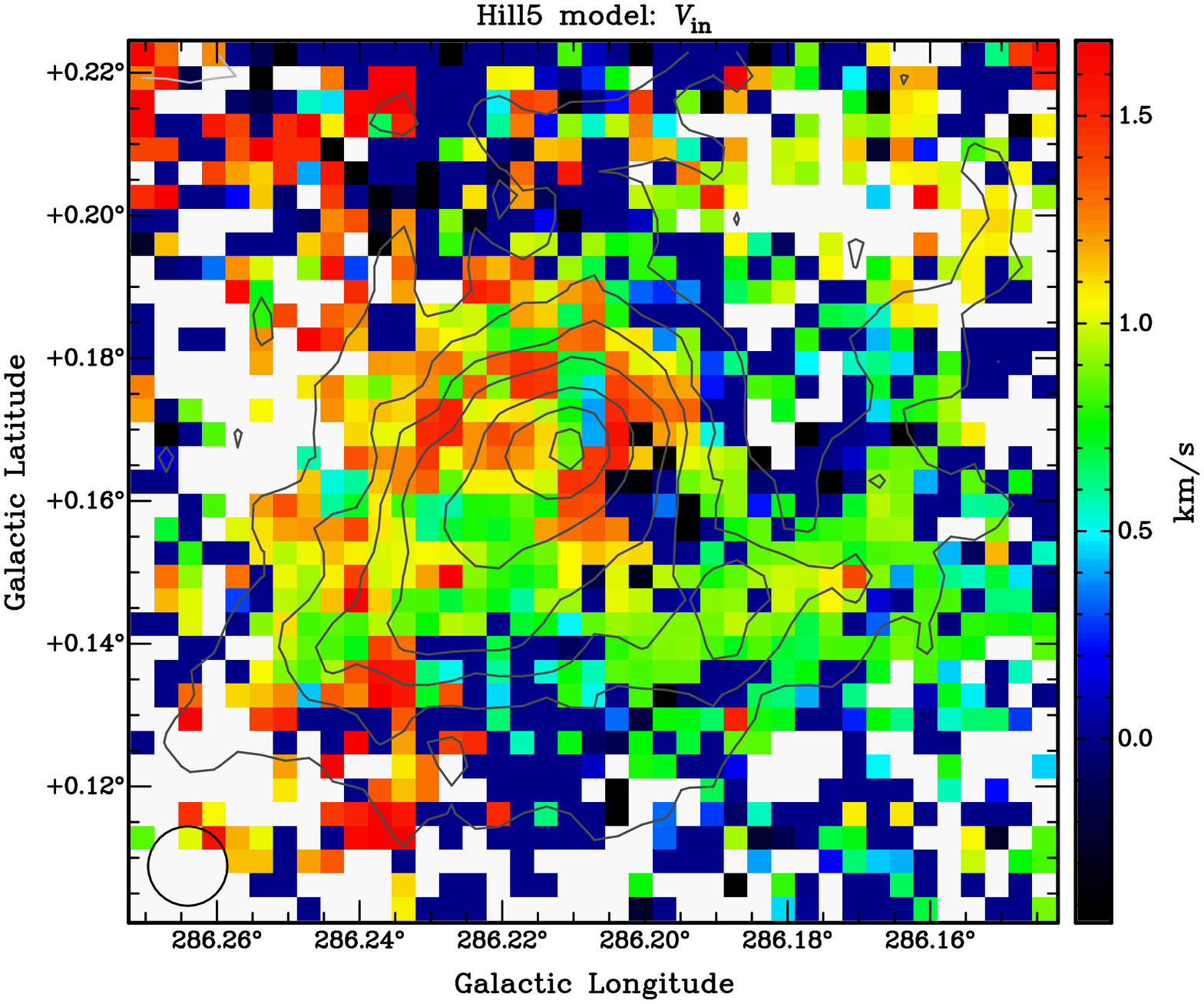}}
\caption{Pixel-by-pixel parameter fits of the Hill5 model to the Mopra \hcop \joz\ cube.  All panels are overlaid with the same \hcop\ integrated intensity contours from Fig.\,\ref{mom1mom2}.  ($a$) Peak excitation temperature.  ($b$) Peak optical depth.  ($c$) Intrinsic $V_{LSR}$.  ($d$) Velocity difference between Fig.\,\ref{mom1mom2}$a$ and panel $c$ in this figure.  ($e$) Intrinsic velocity dispersion.  ($f$) Infall speed.
\label{hill5}}
\end{figure*}

The parameters fitted by the Hill model to an infall spectrum include the peak line excitation temperature and optical depth, the intrinsic (i.e.,\,equivalent optically-thin) $V_{LSR}$ and velocity dispersion, and the infall speed $V_{in}$.  These fits, in order to be considered reliable by \citet{DM05}, should be to spectra with S/N $>$ 30.  Our Mopra \hcop \joz\ cube does not formally satisfy this requirement per 0.11\kms\ channel (peaking at S/N $\sim$\,12), but since the fits are over a large velocity range (up to $\sim$8\kms\ to zero-power) we suggest that the figure of merit should rather be the peak S/N in the integrated intensity map (Fig.\,\ref{mom0}), which is $\sim$\,50.  Put another way, by fitting 5 parameters to spectra with $\sim$70 independent resolution elements, the problem is more strongly constrained than the per-channel S/N would suggest.  Our claim of reliability is bolstered by the solutions themselves, which show little statistical noise in the output parameters above an integrated intensity $\sim$\,1.5\,K\kms, unless the optical depth is too low to give a good infall solution (i.e.,\,everywhere except toward the SW edge of the clump).  In such areas, the fitted $T_{ex}$ and $\tau$ especially are poorly constrained and not physical.

In Figure \ref{hill5} we show the parameter maps of the Hill5 model fits to the \hcop \joz\ line.  The lowest reliable excitation temperature (panel $a$) is $\sim$\,3--4\,K around the northern and eastern perimeter of the clump, and reaches a maximum $\sim$\,6\,K near the peak emission.  These values are clearly less than the $T_{ex}$ discussed below, but this is partly attributable to the data cube being on the $T_A^*$ scale, which is 0.64$\times T_{mb}$.  The only effect this choice of temperature scale has on the model fits is on the scaling of the fitted $T_{ex}$.  Over the same areas, the optical depth (panel $b$) ranges from $\sim$\,1.5 to $\sim$\,6.  The areas where the infall solution is unphysical are clearly visible in these two panels as the red or black pixels, predominantly to the south and/or west of the clump.  In the brighter areas where we believe the solutions are reasonable, we note that the highest optical depth lies NE of the brightest \hcop\ emission, precisely where the brightest \httco\ ridge lies.  Furthermore, the \hcop/\httco\ line ratios (ranging from $\sim$6--10; see Fig.\,\ref{mom0}) are entirely consistent with these optical depths and normal abundance ratios of these molecules.  This is actually remarkable since the Hill5 model only fits parameters to the \hcop\ cube: the fact that the \httco\ data are consistent with the model results gives us further confidence in the Hill5 solutions.  Likewise, the highest $T_{ex}$ is to the SW of the \hcop\ peak, approximately facing the HII region (cf.\,\S\S\ref{distance},\ref{IRfeatures}) and entirely consistent with that geometry.

The various velocity parameters of the Hill5 models are also remarkably well-behaved.  The systemic $V_{LSR}$ map (panel $c$) looks grossly similar to the moment-1 map (Fig.\,\ref{mom1mom2}$b$), but in fact is slightly redshifted where the infall profile is most prominent.  This is made clear in panel $d$ which shows the velocity difference (moment-1)--($V_{LSR}$).  This quantity should be close to zero in most places, but skewed to negative values where the infall is strong and the moment-1 values reflect the blue asymmetry of the spectra.  Indeed the colours in panel $d$ show exactly this: away from the robust infall solutions, the average colour is orange corresponding to a mean difference $\sim$\,0\kms.  Where panels $a$ and $b$ have good solutions, the mean velocity difference is consistently $\sim-0.5$\kms, indicating the extent of the spectral asymmetry.

The last two panels of Figure \ref{hill5} show the velocity dispersion (panel $e$) and infall speed (panel $f$).  In the area of good fits, the former is $\sim$\,1.0$\pm$0.2\kms, while the latter is $\sim$\,1.0$\pm$0.4\kms.  Once again, we see these \hcop-derived dispersions are consistent with the actual linewidth measurements of the \httco\ (\S\ref{mopra}).  Furthermore, as predicted by \citet{DM05} the Hill5 solutions do indeed scale to higher-mass regions than they examined, since for BYF73 the criterion that the intrinsic dispersion is comparable to the infall speed is satisfied.

In panel $f$ we note an interesting structure in the velocity field of the infall.  Along the long axis of the clump (i.e.,\,to the NW and SE), the infall speeds are consistently lower than 1\kms, whereas across the short axis (to the NE and SW) the infall speeds are consistently higher than 1\kms.  It is tempting to interpret this pattern as due to a partially rotationally-supported oblate clump.  In this scenario, the infall is somewhat centrifugally hindered in the equatorial plane (which is roughly parallel to the major axis of the emission) by a rotational speed which may be $\sim$\,0.5--1\kms, but is unimpeded along the presumed rotational axis (roughly parallel to the minor axis).

Equally, we note that the $V_{LSR}$ field in panel $c$ is purely kinematic, since any radiative transfer effects would have been filtered by the model into just $T_{ex}$ and $\tau$ (panels $a$ and $b$).  Thus panel $c$ may be a better indicator of rotation in BYF73, with the rotation axis being roughly aligned with the emission's major axis instead, and suggesting a more prolate geometry for BYF73.  However we also note that the most redshifted portion of panel $c$, centred near (286\degree.195, 0.\degree157), has already been attributed to a separate, non-infalling component in the data cube (\S\ref{mopra}).  Both of these rotation interpretations are thus quite speculative: the noise in both the data and the model may dominate the features we are trying to interpret, and these alternatives really need to be explored with higher resolution and greater sensitivity data in order to discern between them.

An important feature of the Hill5 $V_{in}$ treatment is that it is consistent with the value obtained in \S\ref{gravinfall} based on the \citet{MMT96} work, which was for a two-layer radiative transfer model.  \citet{DM05} similarly found that their two-layer models often gave solutions for $V_{in}$ which were mostly consistent with the Hill models, however the Hill5 model was the most robust to errors.  In the calculations below we use the Hill5 result $V_{in}$ = 1.0$\pm$0.4\,\kms.  For the mass infall rate from \S\ref{gravinfall}, we now have a somewhat larger value
\begin{eqnarray} 
	\frac{dM_k}{dt} & \sim & (3.4\pm1.7)\times10^{-2}\,{\rm M}_{\odot}{\rm yr}^{-1} , 
\end{eqnarray}

\hspace{-4mm}recalling that by evaluating this with the \httco\ radius we are likely obtaining a lower limit to the global infall rate.  We conclude that the radiative transfer modelling of the \hcop\ data gives results which are surprising but highly self-consistent, and consistent with other features of our data.

\subsection{Clump Mass \label{mass}}

Despite the satisfactory results of the modelling, in order to make a strong case for the formation of a massive cluster, we also need to establish that the molecular clump has sufficient mass to qualify for this status, and that other possibilities for interpreting our data are discounted.  Since the gas density will probably be at least the \hcop \joz\ transition's critical density \citep{HW79,BC90} where the bright molecular emission is seen, the cloud mass is given approximately by
\begin{eqnarray} 
	M & > & \mu_{mol} m_{H} n_{cr} (\pi/ln2)^{3/2} R^{3} \\
	 & \sim & 1.0\times 10^4\,{\rm M}_{\odot} \left(\frac{n_{H_{2}}}{3\times10^5\,{\rm cm}^{-3}}\right)\left(\frac{R}{0.40\,{\rm pc}}\right)^3 \nonumber \\ 
	 & \sim & 6.4\times 10^4\,{\rm M}_{\odot} \left(\frac{n_{H_{2}}}{3\times10^5\,{\rm cm}^{-3}}\right)\left(\frac{R}{0.73\,{\rm pc}}\right)^3 \nonumber 
\end{eqnarray}

\hspace{-4mm}using the volume for a 3D Gaussian.  Here we give two values for the mass based on which size we take for the \hcop-emitting region.  With ASTE's detection of the \jft\ line, even higher-density gas ($\sim$10$^7$ cm$^{-3}$) must exist in the clump, and if widespread would give a much higher mass estimate.  Therefore the first value for the mass is almost certainly a lower limit.  However eq.\,(4) assumes that the dense gas giving rise to the emission fills our beam, whereas the filling factor $f$ is unknown and possibly $\ll$1; this may indeed be the case in the outer envelope of the clump, thus the second value is probably an upper limit.

A formally more rigorous, but not necessarily more precise, mass estimate is made (and we obtain an estimate for $f$ as well) if we calculate the \hcop\ column density first. 
We use the full expression without assumptions about optical depth or approximations to the stimulated emission correction in the denominator \cite[e.g.,][]{RW06}.  Assuming LTE applies and with quantities in cgs units, we obtain a column density for {\em each line of sight} from
\begin{mathletters}
\begin{equation} 
	N({\rm HCO^+}) = \frac{3h}{8\pi^3\mu_D^2J_u}\, \frac{ Q(T_{ex})e^{E_u / kT_{ex}}}{(1-e^{-h\nu/kT_{ex}})} \int \tau dV\hspace{1mm} ,
\end{equation}

\hspace{-4mm}where $Q$ is the partition function for \hcop\ at the excitation temperature $T_{ex}$, $E_u$ is the energy level of the upper state $J_u$ of the transition, $\mu_D$ is the molecule's electric dipole moment, and the line optical depth $\tau$ (peak value $\sim$6 from the previous section) is integrated over the velocity, here taken over the range --23.2 to --16.6\,\kms\ (as in Fig.\,\ref{mom0}).

Determining the excitation temperature is a little more complicated, however.  Faundez et al (2004) derive $T_d$ = 30\,K for the continuum dust emission from BYF73, but found it necessary to fit two temperature components to the spectral energy distributions (SEDs) of most of their sources.  They do not give an explicit value for the warm component in BYF73, but their average warm component has $T_d \sim$ 140\,K.  From the \hcop \jft/1$\rightarrow$0 brightness ratio (2.20$\pm$0.14 from Fig.\,\ref{spectra}, suitably corrected for the beam efficiencies) at the peak of BYF73, we fit a $T_{ex}$ = 125$\pm$26\,K for the dense gas.  But without a spatially-resolved \hcop \jft\ map, we are limited to saying that the gas $T_{ex}$ probably takes a range of values from 30--125\,K.  Because \hcop\ is a linear molecule, its partition function is straightforward to calculate \citep{RW06}.  At these temperatures $Q \sim 14-59$, giving 
\begin{eqnarray} 
	N({\rm HCO^+}) & = & 4.84\times10^{11}\, \frac{ Q(T_{ex})e^{E_u / kT_{ex}}}{1-e^{-h\nu/kT_{ex}}} \int \tau dV\hspace{1mm} {\rm cm}^{-2} \nonumber \\ 
	 & \sim & (0.92-13)\times10^{15}\,{\rm cm}^{-2} , 
\end{eqnarray}
\end{mathletters}

\hspace{-4mm}where the velocity is in \kms, and we have taken a Gaussian line profile, with dispersion $\sigma$ as before, for the integral.  Combining the column density with the size measurement (assuming that the physical depth of the source is comparable to the projected size) gives a central density estimate
\begin{eqnarray} 
	n_{H_2} & = & \sqrt{\frac{ln2}{\pi}} \frac{N}{RX} \\
	& \sim & \frac{(3.5-51)\times10^5\,{\rm cm}^{-3}} {(R/0.40\,{\rm pc})(X/10^{-9})} \nonumber 
\end{eqnarray}

\hspace{-4mm}over the same temperature range, which shows that the central density in BYF73 almost certainly exceeds the critical density for thermalising the \hcop \joz\ line, and vindicates this assumption in eq.\,(4).  Similarly integrating the column density (eq.\,5b) over the emission region yields a total cloud mass
\begin{eqnarray} 
	M_{LTE} & = & \frac{N}{X}(\mu_{mol} m_H) \frac{\pi R^2}{ln2} \\
	 & \sim & (1.2-17)\times10^4\,{\rm M}_{\odot} \frac{(R/0.40\,{\rm pc})^2}{(X/10^{-9})} .\nonumber
\end{eqnarray} 

\hspace{-4mm}The lower limits for both eqs.\,(6) \& (7) are probably too low, since they don't include the contribution to the mass from the warmer component; moreover we have used the smaller \httco\ radius in both.  With the larger radius, eq.\,(6) gives a density $\sim$1.9$\times10^5$\,cm$^{-3}$, lower than before but still near the critical density, and eq.\,(7) a mass $\sim$4.1$\times$10$^4$\,M\solar.  Both low-temperature mass estimates are close to the values in eq.\,(4).  This is surprising given the approximate nature of these calculations, but reassuring.  We conclude that an intermediate value, $M \sim$ 2.0$\times$10$^4$\,M\solar, is probably reasonable for BYF73.

However the upper limits from eqs.\,(5-7), based on the higher excitation temperature being widespread, are certainly too high, since it is unlikely that such a warm temperature would be typical of the whole parsec-wide dense clump.  One would need to compare a map of the \hcop \jft\ emission at a resolution at least as good as our \hcop \joz\ map, in order to derive a reliable map of $T_{ex}$ across the source and address how much larger the clump's mass might be due to this warmer gas.

In eq.\,(6-7) we have used an abundance $X = 10^{-9}$ for \hcop\ relative to H$_2$, which is a strong upper limit from some recent models of massive core chemistry \cite[e.g.,][]{GWH08}.  These models show $X_{{\rm HCO}^+}$ is a strong function of time, and is not necessarily the main charge carrier in such regions.  Thus $X_{{\rm HCO}^+}$ in massive cores may be an order of magnitude lower than a more typical value $\sim10^{-9}$ in low-mass cores \cite[e.g.,][]{LWW90,CWZ02, LES03} and used here.  On the other hand, \citet{ZCP09} obtain $X_{{\rm HCO}^+}\sim$ 2.3--12$\times$10$^{-9}$\,cm$^{-3}$ from observations of a sample of massive clumps, although they cautioned that their values are probably overestimates.  This means that parameters derived here that depend on $X$ are quite uncertain.  Nevertheless, eqs.\,(4) and (7) suggest that $f$ may be close to unity, and that $M$ is large.

Indeed BYF73 seems to be quite extreme in this regard too.  For example, the mass surface density corresponding to the column density from eq.\,(5b) is $\Sigma = (N/X)(\mu_{mol} m_H)\sim$ 35\,kg\,m$^{-2}$, which is near the largest value of the massive Galactic clusters considered by \citet{MT03}.  Therefore BYF73 is interesting as a likely environment in which massive protostellar cores may form, and then form massive stars.

Are there alternatives for the dynamical state of this clump besides gravitational infall?  To answer this, we evaluate a number of terms from the Virial Theorem.  If the linewidths seen in the \hcop\ ($\sim$2\kms\ relative to the line centre, counting emission out to the half-power level) were due to rotational support against self-gravity (an interpretation we do not favour due to the self-consistency of the infall modelling, and the effective limit of $\sim$\,1\kms\ to any rotational contribution to the spectral lines), then
\begin{mathletters}
\begin{eqnarray} 
	M_{rot} & = & v^2 R/G \\
	 & \sim & 370\,{\rm M}_{\odot} \left(\frac{v}{2\,{\rm km s}^{-1}}\right)^2 \left(\frac{R}{0.40\,{\rm pc}}\right) . \nonumber 
\end{eqnarray}

\hspace{-4mm}However thermal and magnetic pressure must also contribute to the support of the cloud; the corresponding virial relations give
\begin{eqnarray} 
          M_{th} & = & 5kT_{ex}R/m_{H_2}G \\
           & \sim & 240\,{\rm M}_{\odot} \left(\frac{T_{ex}}{125\,{\rm K}}\right)\left(\frac{R}{0.40\,{\rm pc}}\right) \nonumber 
\end{eqnarray}
and (in cgs units only)
\begin{eqnarray} 
	M_{mag} & = & (5B^2 R^4/18G)^{1/2} \\
	 & \sim & 940\,{\rm M}_{\odot} \left(\frac{2B_{los}}{600\,\mu{\rm G}}\right)\left(\frac{R}{0.40\,{\rm pc}}\right)^2 , \nonumber 
\end{eqnarray}
\end{mathletters}

\hspace{-4mm}where we have taken an appropriate value from studies in similar regions for the magnetic field, up to twice the typical Zeeman-derived $B_{los}$ at a density 3$\times$10$^5$cm$^{-3}$ \cite[see Fig.\,1 of ][]{Cru99}.  Such a large value is further supported by \citet{FTC08} who obtained $B$(median) = 560$\mu$G from CN observations of clouds with a mean density 4$\times$10$^5$cm$^{-3}$.  These terms, even in combination ($\sim$1550 M\solar), are much too small to provide the necessary support against gravity, unless (for example) the magnetic field strength were at least ten times the value assumed here, and/or we take linewidths out to the zero-power level ($\pm$4\,\kms).  While such values for rotation and the magnetic field are not entirely ruled out as a means of supporting BYF73 against collapse, they would be quite extreme.  We infer that virial equilibrium does not apply in this case, despite the \hcop\ abundance and $T_{ex}$ uncertainties.

\notetoeditor{Figures 7a and 7b should appear side-by-side in print, and MUST be in colour.}
\begin{figure*}[t]
(a)\includegraphics[angle=-90,scale=0.34]{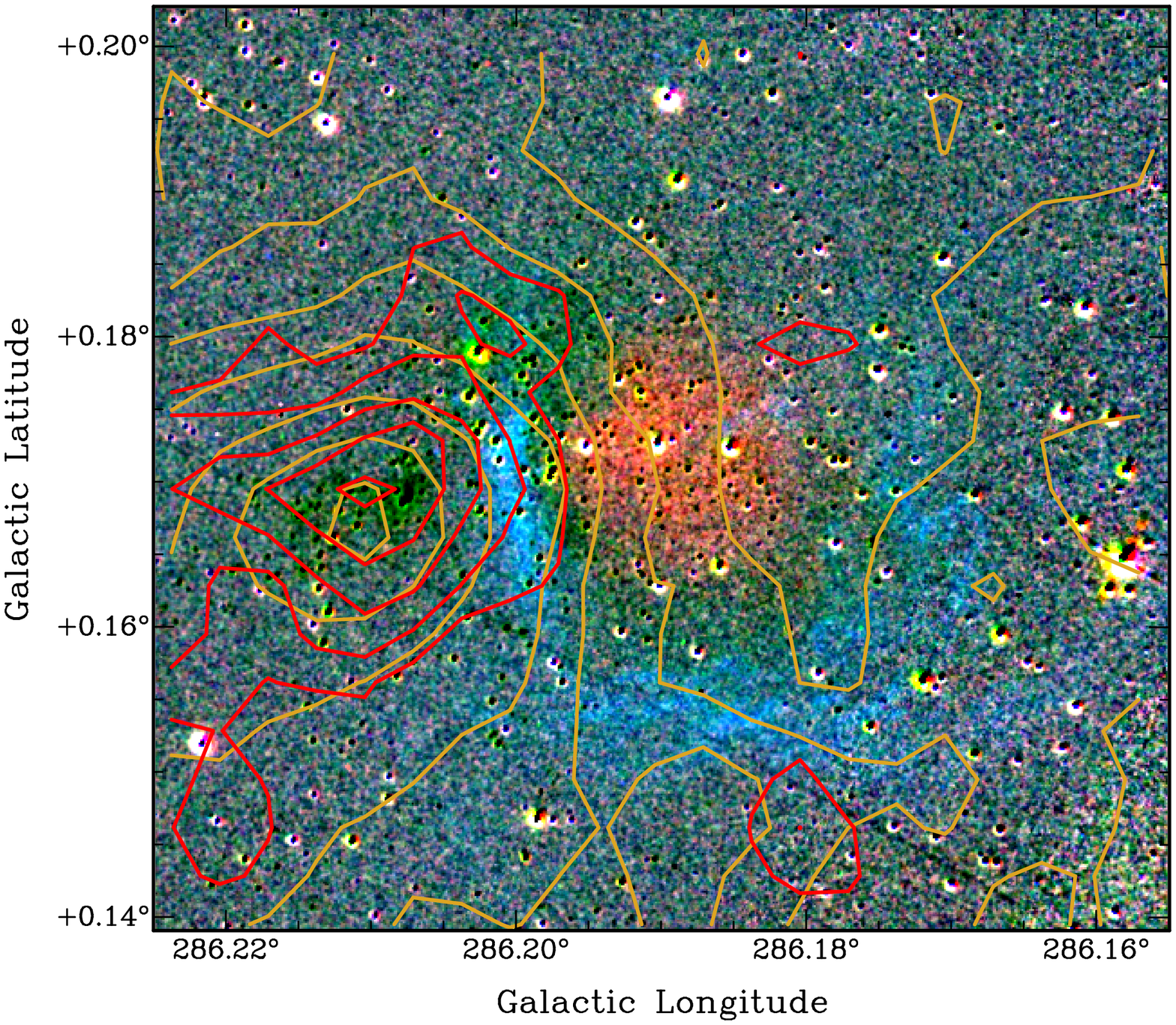}
(b)\includegraphics[angle=-90,scale=0.34]{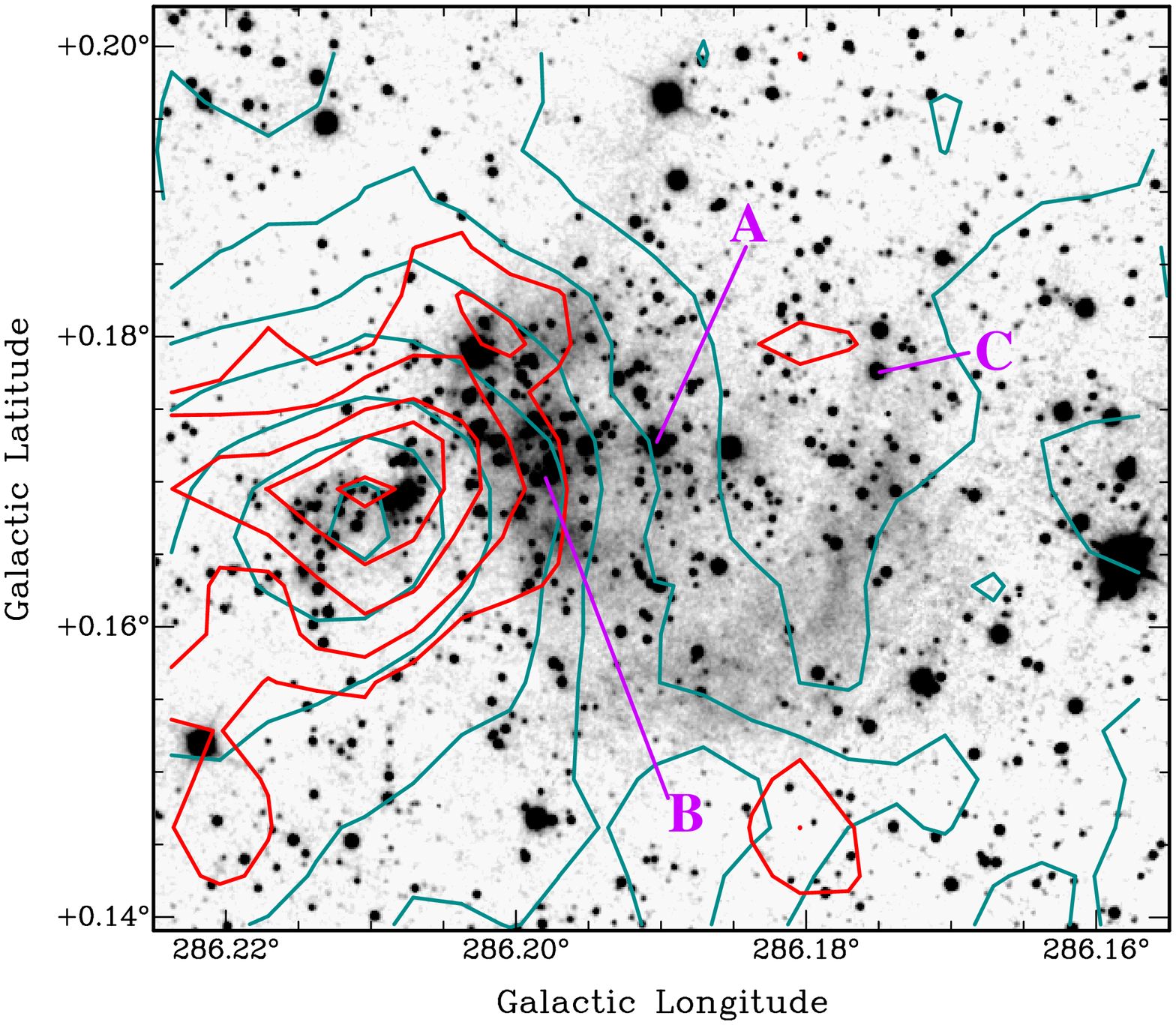}
\caption{(a) RGB-pseudocolour image of BYF73 in $K$-band spectral lines.  Here Br-$\gamma$ is shown as red, and H$_2$ $S$(1) is shown as green ($v$=1$\rightarrow$0) \& blue ($v$=2$\rightarrow$1).  Contours are overlaid from Mopra \hcop\ ({\it gold}) and \httco\ ({\it red}) integrated intensities (levels as in Figs.\,\ref{mom1mom2} \& \ref{mom0}, resp.).  (b) Same contour levels as {\it a}, but now coloured {\it blue} and {\it red}, and overlaid on the $K$-continuum image.  The stars labelled in magenta are the same as those similarly labelled in Fig.\,\ref{IRspec}a.  Recall from Fig.\,1 that at a distance of 2.5\,kpc, the scale is 40$''$ = 0.485\,pc or 0\degree.02 = 0.873\,pc or 1\,pc = 0\degree.0229 = 82$''$\hspace{-1mm}.5.
\label{IRcolour}}
\end{figure*}

There is also the possibility that the velocity pattern in BYF73 represents a massive outflow rather than infall.  Besides the detailed spectroscopic arguments for infall, we discount the outflow interpretation since maps of the \hcop\ line wings (not shown here) do not reveal any particular geometric pattern, such as a bipolar separation of the line wings.  Nevertheless, sensitive $^{12}$CO observations should be made of BYF73, since they might be better able to find any outflow, if present.

The conclusion that BYF73 is indeed a massive dense clump undergoing contraction at least (if not collapse) seems fairly reliable, the strongest evidence being the line profiles, the mass calculations, and the IR appearance (see \S\ref{IRfeatures}).  We can compare our mass estimate for BYF73 with others in the literature.  From Nanten CO mapping and IRAS fluxes, \citet{YAK05} obtained LTE and virial cloud masses of 1900 and 3600\,M\solar\ (resp.)\ and a luminosity 3.0$\times$10$^4$\,L\solar.  Our LTE mass estimates are significantly larger than theirs, likely due to our use of a tracer of denser gas, but our total virial mass is smaller than theirs, as might be expected with a smaller observed linear size and linewidth.  Our LTE estimates would be smaller if we assumed a smaller effective density for \hcop\ and/or a larger \hcop\ abundance.  Either of these might reduce our best estimate above by a factor of 3 or so, to $<$10$^4$\,M\solar, bringing it more into line with the \citet{YAK05} number, although we do not favour this value.  From 1.2mm mapping and SED fitting \citet{FBG04} obtained a clump mass of 470\,M\solar, density $n_{H_2}$ = 1.4$\times$10$^5$\,cm$^{-3}$, dust temperature 30\,K, and luminosity 1.9$\times$10$^4$\,L\solar.  Their mass value seems quite low, but could be as much as five times higher with a lower assumed dust opacity, as they point out.  This would bring it into closer agreement with the \citet{YAK05} mass estimate, and suggests that such lower dust opacities may be required to explain the higher molecular masses.  At any rate, our \hcop\ data strongly suggest the presence of a large amount of dense gas that may not be fully sensed in CO lines or in the mm-continuum.

\notetoeditor{Figures 8a and 8b should appear side-by-side in print.}
\begin{figure*}[t]
\hspace{-4mm}\vspace{-5mm}
\includegraphics[angle=-90,scale=0.34]{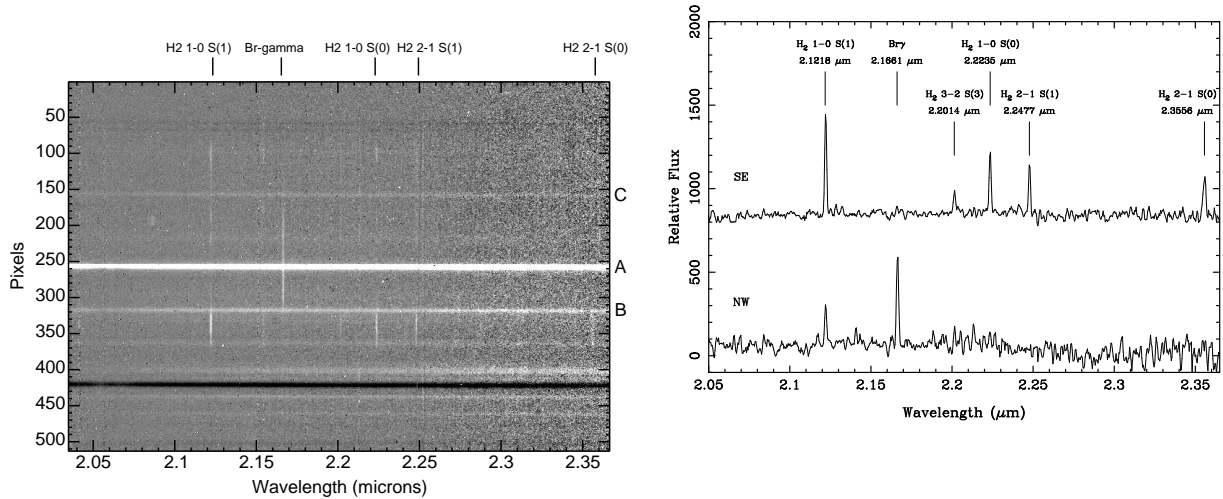}
\includegraphics[angle=-90,scale=0.31]{figures/fig8b.ps}
\caption{(left) Near-IR long-slit spectrum across BYF73, where the slit was oriented along a line joining the peak of the HII/Br-$\gamma$ emission (to the NW of Fig.\,\ref{IRcolour} at the top of this figure) and the peak of the \hcop\ emission (to the SE of Fig.\,\ref{IRcolour} at the bottom of this figure).  The spectral lines are as indicated, and the labels A, B, and C refer to stars similarly labelled in Fig.\,\ref{IRcolour}.  (right) Sample spectra from panel (a) on a relative flux scale to indicate line ratios.  The upper spectrum (labelled ``SE'') corresponds approximately to pixel row 340 in panel (a), while the lower spectrum (labelled ``NW'') is near pixel row 200.
\label{IRspec}}
\end{figure*}

\subsection{Infrared Features \label{IRfeatures}}

Our $K$-band imaging (Fig.\,\ref{IRcolour}) has much higher angular resolution than our mm data, or archival centimeter-wave (cm) and far-infrared (FIR) data, and shows some striking structures and correlations.  Near the molecular clump, there is a compact HII region (visible as a Br-$\gamma$ emission nebula) and IR cluster (visible also in the $K$-continuum image).  The Br-$\gamma$ is exactly coincident with a centimetre-continuum point source from both the Molonglo Galactic Plane Survey-2 \cite[the MGPS-2 has a similar beamsize, $45''\times53''$, to our Mopra data;][]{MMG07} and the slightly lower-resolution Southern Galactic Plane Survey \citep{HGM06}.  Such features have been seen before around similar massive star-forming dense clumps, e.g.\,NGC 2024 \citep{BCB89} or AFGL 5179 \citep{TOG06}, but the example of BYF73 is interesting in the rather clean separation of the ionised and molecular components, and the distinct ``cocooning'' of the excited H$_2$ emission around the very symmetric Br-$\gamma$ and cm-continuum.  Indeed, this is actually reminiscent of planetary nebulae \cite[e.g., Fig.\,9 of][]{RSA98} or the classic picture of a Str\"omgren sphere.  From the pseudocolour composite image in Fig.\,\ref{IRcolour}a, we see that the shell of excited H$_2$ appears to be traced much better by the $v$=2$\rightarrow$1 than the $v$=1$\rightarrow$0 emission.  This is surprising since a [1--0]/[2--1] ratio less than unity would be at odds with our understanding of H$_2$ excitation.  Instead, this ratio is likely either an artifact of differential reddening between the various filters used, or due to non-photometric imaging conditions, or both.

To confirm this we obtained a $K$-band long-slit spectrum aligned between the mm and Br-$\gamma$ peaks (Fig.\,\ref{IRspec}), which shows that the [1--0]/[2--1] ratio is actually 2.14$\pm$0.10 at the molecular-ionised interface, typical of photodissociation regions (PDRs) or shock-excited jets \cite[e.g.,][]{AJL05,CGN06}.  From the measured $S$(1) [1--0]/[2--1] ratio and the tabulation of T.\,Geballe (1995)\footnote{ Quoted by T.\,Kerr (2004), www.jach.hawaii.edu/UKIRT/ astronomy/calib/spec\_cal/h2\_s.html}, the gas kinetic temperature at the PDR interface is constrained to be $>$4000\,K; including the $S$(0) and $S$(3) lines also visible in Figure \ref{IRspec} (with respective ratios $S$(1) 1--0 to these lines of 1.65$\pm$0.07 and 4.5$\pm$0.3) suggests a temperature as high as $\sim$5000\,K.  This is comparable to, but somewhat higher than, H$_2$ temperatures seen in other star formation PDRs \citep{AJL05} or low-mass H$_2$ jets \citep{CGN06}, approaching the typical $T_e \sim$ 7000\,K for Galactic HII regions at this galactocentric radius \cite[8\,kpc;][]{SMN83}, and is perhaps indicative of the relative youth of the HII region in BYF73, and/or the strength of the shock excitation from the young stars in the HII region.

Another remarkable feature of the IR imaging is the apparent deficit in line emission from the exact peak of the mm-molecular emission (within the pointing uncertainty).  Unfortunately a spectrum for this position is not available, since it coincides with emission in the reference beam (the horizontal black line in Fig.\,\ref{IRspec} at pixel coordinate 420).  So while we cannot obtain any IR line ratios here with the current data, it appears as if the H$_2$ $v$=2$\rightarrow1$ and Br-$\gamma$ are both seen in absorption at this peak, creating an apparent ``absorption nebula''.  This nebula can be seen in Figure \ref{IRcolour}a as a green patch to the left (Galactic east) of the HII region, since there the blue (H$_2$ $v$=2$\rightarrow$1) and red (Br-$\gamma$) appear more strongly ``absorbed'', while the green (H$_2$ $v$=1$\rightarrow$0) is only weakly ``absorbed''. While much of this appearance may be due to the construction of the RGB image, at the very least it is likely that there is either unusual, highly localised IR emission/absorption at the molecular peak, or that the deeply embedded stars at that position have very unusual IR colours.  Moreover, this positional coincidence is highly suggestive.  Further east of the peak of the absorption nebula, there seems to be a weaker, comma-shaped extension, as well as a highly reddened cluster of stars; this shape is also seen in some of the \hcop\ channel maps.  Furthermore, this extension is also aligned with the supposed equatorial plane of the oblate clump scenario from \S\ref{radxfer}.

There is evidence in other archival data for an unusual source at the molecular peak.  A 3-colour (i.e.,\,$JHK$) 2MASS image shows that the HII region exhibits some moderate reddening, but that the knot of $K$-band emission visible in Figure \ref{IRcolour}$b$ at the molecular peak is virtually invisible at shorter wavelengths, confirming its highly embedded nature.  A similar 3-colour MSX image (i.e.,\,8,15,21\,$\mu$m) further shows that this embedded source dominates the luminosity at MIR and longer wavelengths.  We therefore have the rather unusual situation that, while a more evolved source is adjacent to our molecular clump, the apparently less evolved source(s) within the clump are more luminous than the revealed exciting stars.  Indeed, the luminosity of the deeply embedded IR source(s) is at least partly derived from the release of gravitational potential energy.  Taking the mass inflow rate from \S\ref{radxfer}, which is itself probably a lower limit as described there,
\begin{eqnarray} 
	L_{grav} & = & \frac{GM\dot{M}} {R} \\
	 & \sim & 1200\,{\rm L}_{\odot} \frac{(M/20,000\,{\rm M}_{\odot}) (\dot{M}/0.034\,{\rm M}_{\odot}{\rm yr}^{-1})} {(R/0.40\,{\rm pc})} . \nonumber 
\end{eqnarray}

\hspace{-4mm}This result, that $L_{grav}\gapp 4\%L_{bol}$, may be even larger if higher-resolution observations of the infall and central MIR/NIR sources reveal the inflow continues deeper into the central regions.

If the NIR ``absorption'' at this position were real and not an imaging artifact, then it would imply very large columns of gas, $\sim$10$^{24}$\,cm$^{-2}$.  The \hcop\ self-absorption and spatial distribution require the same condition, and so the implication of very high column density at the molecular peak would seem to be strong.  This is further supported when we note that the continuum emission of the three stars closest to the ``absorption nebula'' position in Figure \ref{IRspec}a (i.e.,\,the bright horizontal lines at pixels 400, 435, and 460) show a very strong attenuation at the shorter $K$-band wavelengths, presumably due to severe reddening in the molecular clump.  Such reddening is not apparently affecting the stars in the HII region (e.g., those labelled A, B, \& C in Fig.\,\ref{IRspec}) to the same degree.

The $K$-continuum image reveals details of clustering in BYF73.  Compared to the surrounding sky away from any \hcop\ emission, within the HII region there is clearly an overabundance of stars.  In addition, the bright H$_2$ nebulosity immediately to the east of the HII region contains an even more compact clustering of brighter stars, and there is another tight grouping around the molecular peak.  To the north and south of the molecular peak, the star density is actually lower than the surroundings, suggesting that here the dust column density is still so high that background stars are being extinguished at 2$\mu$m.  It is clear that many young stars, some massive enough to form an HII region, have already formed in BYF73, and that further star formation appears to be proceeding vigorously to the east of this HII region.

\subsection{Theoretical Considerations \label{theory}}

We note that the typical projected nearest-neighbour separations of stars in these IR groups (i.e.,\,at the molecular and Br-$\gamma$ peaks, and molecular-ionised interface), $\sim$2$''$ or 5000 AU by inspection of Figure \ref{IRcolour}b, is less than the Jeans length
\begin{eqnarray} 
	R_{Jeans} & \sim & \left(\frac{kT_{kin}}{G(\mu_{mol} m_H)^2 n_{H_2}}\right)^{0.5} \\
	 & \sim & 7900\,{\rm AU} \left(\frac{T_{kin}}{30\,{\rm K}}\right)^{0.5} \left(\frac{n_{H_2}}{3\times10^5\,{\rm cm}^{-3}}\right)^{-0.5} \nonumber 
\end{eqnarray}

\hspace{-4mm}in the dense clump, and although higher densities, especially near the centre, may make these scales more commensurate, our estimate for eq.\,(10) is probably a strong lower limit considering that there is warmer gas in the clump (\S\ref{mass}), and that the 1mm-derived density \cite[1.4$\times$10$^5$\,cm$^{-3}$, ][]{FBG04} is lower than that used above.  This disparity is typical of massive clusters \cite[e.g.][]{Chu02} and is a well-known feature of such regions that models must reproduce.  Currently, theories attempt to model this structure using either competitive processes \cite[such as coalescence, e.g.][]{BBV03} or a scaled up accretion disk/turbulent core scenario \cite[e.g.,][]{MT03}.  BYF73 promises to be a useful test case in this debate, but as suggested by the IR imagery, will require mm-interferometric observations that approach the IR resolution.  At this level ($\sim$1$''$ or better) we begin to match (at the distance of BYF73) the spatial resolution ($\sim$0.01\,pc) in the simulations of \citet{BP07}.  High-resolution maps of the gas velocity field and linewidth will then help to discriminate between the competing theories.

From the current mm data we can say that the line emission pattern and derived velocity field in BYF73 are consistent with the detailed MHD simulations of \citet{BP07} and radiative transfer treatment of \citet{ZEK93} for protostars, as well as with the treatment of \citet{MT03}.  However in this case (a) the mass infall rate and mass \& size scales are much larger than in any of these models, (b) there are multiple protostellar objects within the collapse zone, rather than a single, more massive one, and (c) the canonical spectral energy distribution of low-mass Class 0 protostars \citep{AWB00} has the flux dropping to undetectable levels at wavelengths shortward of 10$\mu$m, although this is under the assumption of spherical symmetry.  In BYF73 there are a number of near-IR sources visible at the centre of the infalling clump, so it is likely that orientation effects play a role in the emergent SED for massive protostars, and/or that the SED evolution is different in the massive protostar case.  Again, higher-resolution mm- and FIR-continuum images of the cluster sources will help delineate SED evolution in these massive protostars.

Furthermore, this ongoing star formation is happening within a large-scale infall region ($\sim$1\,pc), and to our knowledge this is the first time that such a coincidence of phenomena has been seen.  With so much gas still infalling, it is possible that BYF73 could form many more stars before the supply of material is consumed.  Dividing the clump's mass by the infall rate gives a maximum lifetime $t \sim 6\times10^5$\,yr for the supply of raw material for new stars, although if a protostar massive enough to develop its own HII region ionises the gas and arrests the infall, the cluster's formation may be complete in much less time.  This is very long compared to a dynamical timescale, $t_{ff}=\pi(R^3/8GM)^{1/2}\sim 30,000$\,yr, and tends to support longer-timescale models such as the \citet{TKM06} ``Equilibrium Cluster Formation'' model, rather than (e.g.) the rapid star cluster formation models of \citet{Elm00,Elm07}.  However the most embedded stars in the IR ``absorption nebula'' seem to be arranged in a filamentary geometry; being non-symmetric, such distributions tend to support a rapid formation scenario, however this appearance may be affected by the strong extinction in the area.

In summary, we claim that the molecular and IR observations of BYF73 indicate the existence of a dense clump undergoing global gravitational infall, similar to the case for NGC 2264C \citep{PAB06}.  We further suggest that, on the basis of the size, mass, luminosity, rate of infall, and near-IR appearance, BYF73 is in the process of forming a massive protocluster.  The global mass infall rate as determined from the Mopra mm observations is very high even for a massive protostar, $(3.4\pm1.7)\times10^{-2}$\,M\solar\,yr$^{-1}$ or more.  To our knowledge, the upper end of this range would be unprecedented, if confirmed.

\section{Conclusions}

From Mopra and ASTE HCO$^+$ observations, the Galactic source G286.21+0.17 (which we also call BYF73 from the CHaMP survey master list) has been found to be a massive dense molecular clump exhibiting clear signs of gravitational infall.  The size and scale of this infall, $dM/dt \sim 3.4\times10^{-2}$\,M\solar\,yr$^{-1}$ over $\sim$1\,pc, is either a record or close to it, and may indicate the global formation of a massive protocluster.  AAT near-IR imaging confirms the existence of unusual spectral signatures and a deeply embedded cluster of stars in the infall zone, as well as an adjacent compact HII region and young star cluster.  Higher-resolution mm-wave and FIR/MIR observations of this source are encouraged, since it appears to be an exemplary test case for confronting competing theories of massive star formation.


\acknowledgments

Users of the Mopra telescope have benefited immensely from the efforts of many people over the last several years, including the talented and dedicated engineers and scientists at the ATNF, and the staff and students of the ÒStar FormersÓ group within the School of Physics at the University of New South Wales.  Because of these efforts, use of this facility has changed over this period from being a difficult exercise to a real pleasure.  We would also like to acknowledge the members of the ASTE team for the operation and ceaseless efforts to improve ASTE.  This work was financially supported in part by Grant-in-Aid for Scientific Research (KAKENHI) on Priority Areas from the Ministry of Education, Culture, Sports, Science, and Technology of Japan (MEXT), No. 15071205.

PJB gratefully acknowledges support from the School of Physics at the University of Sydney and through NSF grant AST-0645412 at the University of Florida.  AMH acknowledges support provided by the Australian Research Council (ARC) in the form of a QEII Fellowship (DP0557850).  YM acknowledges financial supports by the research promotion scholarship from Nagoya University and research assistantships from the 21st Century COE Program ``ORIUM'' (The Origin of the Universe and Matter: Physical Elucidation of the Cosmic History) and the Global COE program ``Quest for Fundamental Principles in the Universe: from Particles to the Solar System and the Cosmos'', MEXT, Japan.

We also thank the referee for numerous helpful suggestions and comments which led to several improvements in the paper, and J. Tan and C. De Vries for additional helpful comments.


{\it Facilities:} \facility{Mopra (MOPS)}, \facility{AAT (IRIS2)}, \facility{ASTE}




\clearpage


\end{document}